\documentclass[journal]{IEEEtran}
\usepackage{graphicx}
\usepackage{color}
\usepackage[colorlinks,
			linkcolor=red,
			anchorcolor=blue,
			citecolor=green
			]{hyperref}
\usepackage{bm}
\usepackage{multirow}
\usepackage{cite}
\usepackage{amsmath}
\usepackage{amssymb}
\usepackage{array}
\usepackage{etoolbox,siunitx}
\usepackage[normalem]{ulem}
\usepackage[subrefformat=parens,caption=false,font=footnotesize]{subfig}
\usepackage{algorithm}
\usepackage{algorithmic}

\newcommand{\multirowcell}[1]{\begin{tabular}[b]{@{}c@{}}#1\end{tabular}}
\robustify\bfseries
\robustify\uline

\begin{document}

\title{Visually-Guided Sound Source Separation with Audio-Visual Predictive Coding}

\author{Zengjie~Song and Zhaoxiang~Zhang,~\IEEEmembership{Senior Member,~IEEE}%
\thanks{Manuscript first version received May 10, 2021; second version received February 22, 2022; third version received November 28, 2022; revised April 12, 2023; accepted June 12, 2023. This work was supported in part by the Major Project for New Generation of AI under Grant 2018AAA0100400; in part by the National Natural Science Foundation of China under Grant 61836014, Grant U21B2042, Grant 62072457, Grant 62006231, and Grant 61976174; and in part by the Project funded by China Postdoctoral Science Foundation under Grant 2021M703489. \emph{(Corresponding author: Zhaoxiang~Zhang.)}
	
	Zengjie~Song is with the School of Mathematics and Statistics, Xi’an Jiaotong University, Xi’an 710049, China (e-mail: zjsong@hotmail.com).
	
	Zhaoxiang~Zhang is with the Center for Research on Intelligent Perception and Computing, National Laboratory of Pattern Recognition, Institute of Automation, Chinese Academy of Sciences, Beijing 100190, China, and also with the Center for Artificial Intelligence and Robotics, Hong Kong Institute of Science \& Innovation, Chinese Academy of Sciences, Hong Kong, China (e-mail: zhaoxiang.zhang@ia.ac.cn).
	}
}

\markboth{Journal of \LaTeX\ Class Files,~Vol.~6, No.~18, June~2023}%
{Song \MakeLowercase{\textit{et al.}}: Visually-Guided Sound Source Separation with Audio-Visual Predictive Coding}

\maketitle

\begin{abstract}
	The framework of visually-guided sound source separation generally consists of three parts: visual feature extraction, multimodal feature fusion, and sound signal processing. An ongoing trend in this field has been to tailor involved visual feature extractor for informative visual guidance and separately devise module for feature fusion, while utilizing U-Net by default for sound analysis. However, such divide-and-conquer paradigm is parameter inefficient and, meanwhile, may obtain suboptimal performance as jointly optimizing and harmonizing various model components is challengeable. By contrast, this paper presents a novel approach, dubbed audio-visual predictive coding (AVPC), to tackle this task in a parameter efficient and more effective manner. The network of AVPC features a simple ResNet-based video analysis network for deriving semantic visual features, and a predictive coding-based sound separation network that can extract audio features, fuse multimodal information, and predict sound separation masks in the same architecture. By iteratively minimizing the prediction error between features, AVPC integrates audio and visual information recursively, leading to progressively improved performance. In addition, we develop a valid self-supervised learning strategy for AVPC via co-predicting two audio-visual representations of the same sound source. Extensive evaluations demonstrate that AVPC outperforms several baselines in separating musical instrument sounds, while reducing the model size significantly. Code is available at: \href{https://github.com/zjsong/Audio-Visual-Predictive-Coding}{https://github.com/zjsong/Audio-Visual-Predictive-Coding}.
\end{abstract}


\section{Introduction}\label{sec:intro}
\IEEEPARstart{S}{urrounded} by diverse objects and receiving multisensory stimuli, the human brain conducts multimodal perception to understand the physical world. In fact, the inherent correlations rooted in co-occurred sensory modalities have the potential to facilitate decision making on individual tasks \cite{Hartmann33,Neil06,Stein08,Koelewijn10}. One practical example emerges from the interplay between auditory and visual senses, namely in addition to by hearing the sound of a concert, audiences could be got in the mood for melodious music much easier by watching musicians' body and hand movements. With such inspiration in mind, researchers have been making great efforts to explore the interaction of vision and audio information in multimodal learning. Representative studies include sound recognition \cite{Aytar16,Arandjelovic17,Korbar18}, cross-model retrieval \cite{Arandjelovic18,Suris18,Gabeur20,Chen21a} and generation \cite{Hao18,Zhou18,Wu22}, sound localization \cite{Hershey99,Senocak18,Hu19,Qian20,Chen21b,Song22}, sound source separation \cite{Owens18,Gao18,Zhao18,Gao19,Xu19,Zhao19,Gan20,Zhu20a,Chatterjee21,Chatterjee22}, \emph{etc}. For concreteness this work focuses on the visually-guided sound source separation (also referred as audio-visual sound separation or visual sound separation), which aims to recover sound components from a mixture audio with the aid of visual cues. As one of the fundamental sound processing tasks, visual sound separation promotes a wide range of downstream applications, such as audio denosising, dialog following, audio event remixing, audio-visual video indexing, video sound editing, instrument equalization, and embodied navigation \cite{Zhao18,Gao19,Wang22}.

\begin{figure}
	\centering
	\subfloat[\label{fig:compare_framework_previous}]{\includegraphics[width=0.8\linewidth]{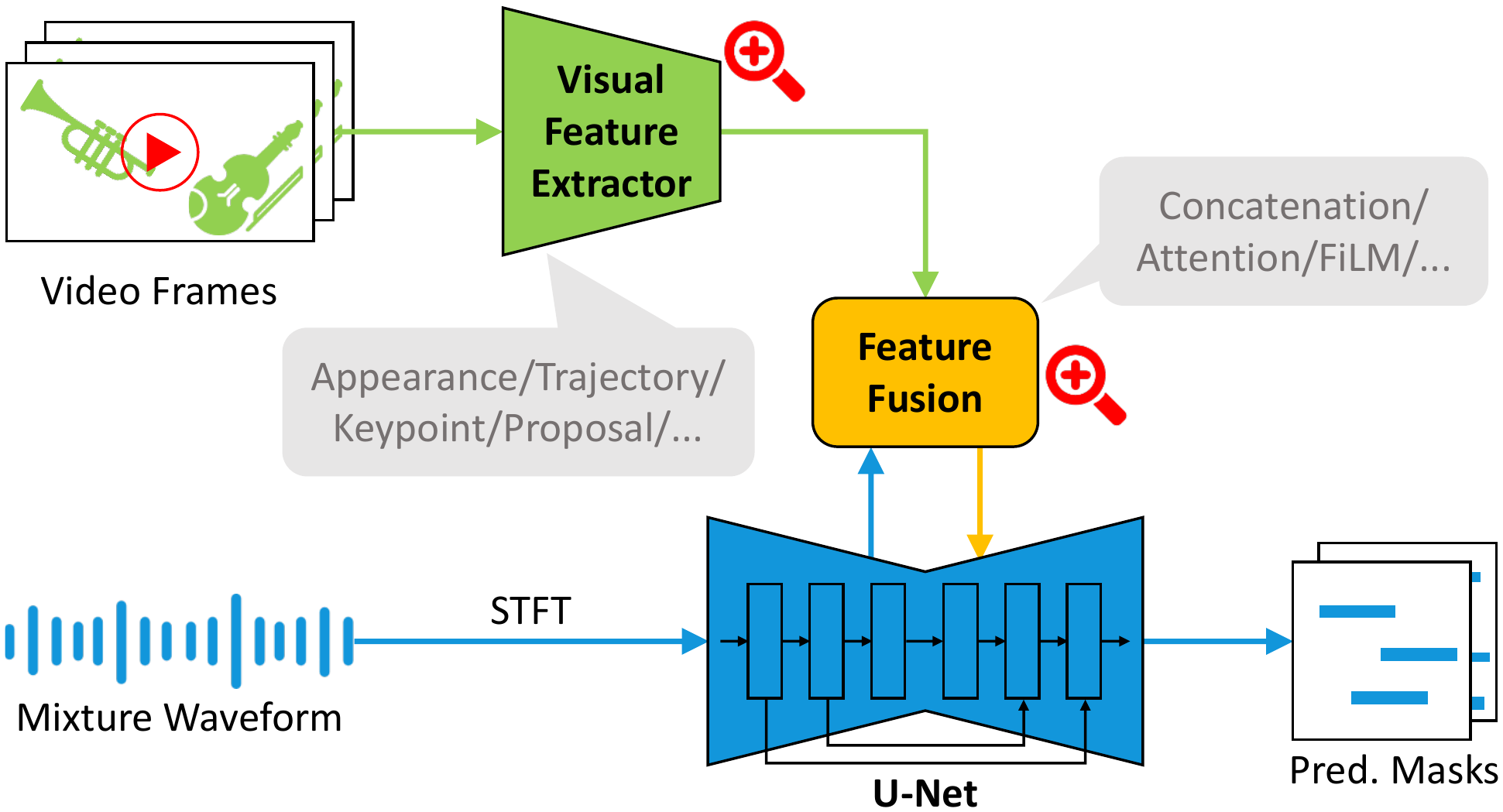}}\\
	\subfloat[\label{fig:compare_framework_ours}]{\includegraphics[width=0.8\linewidth]{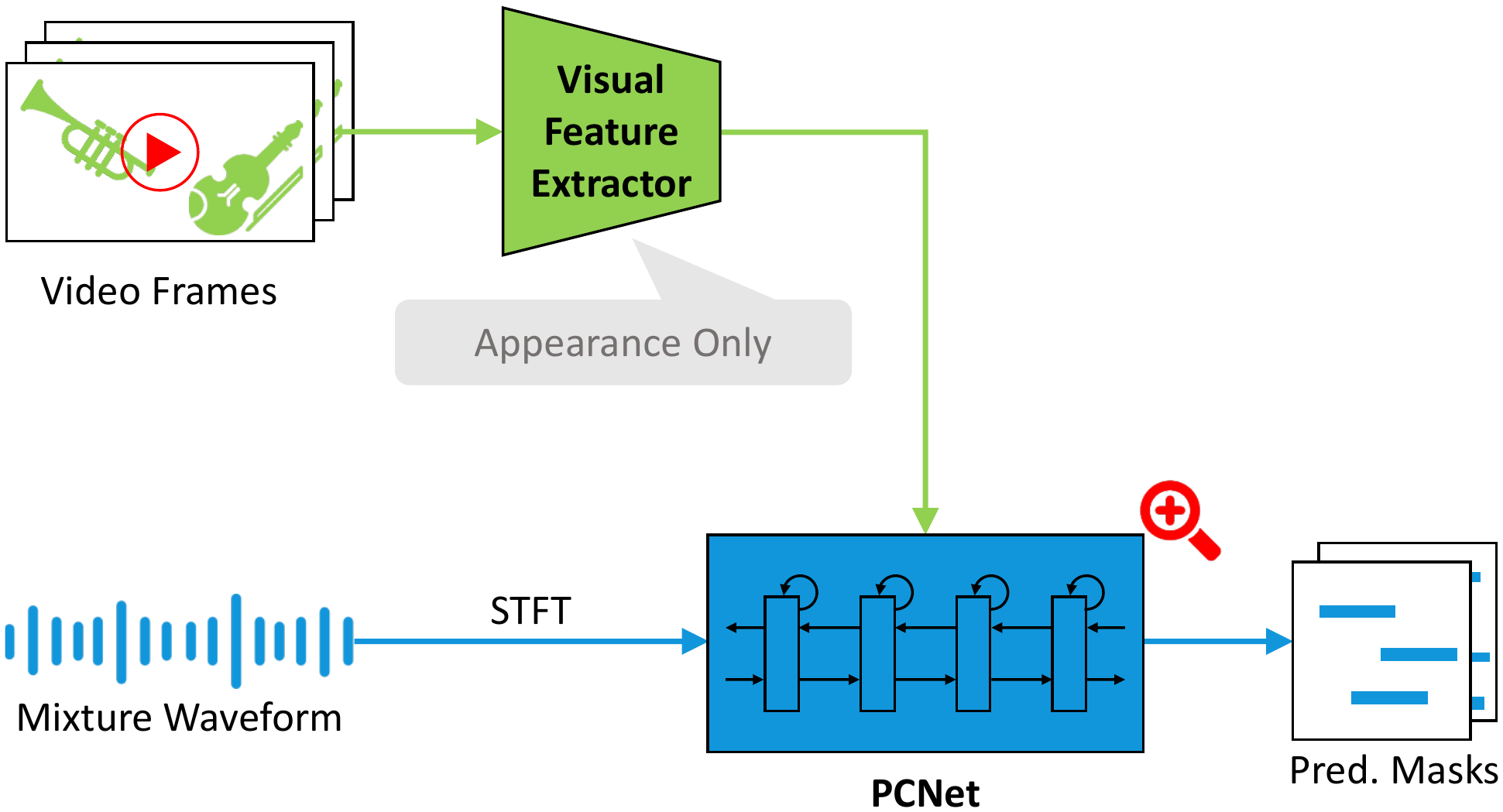}}
	\caption{Comparison with previous methods and our approach on visual sound separation. (a) Previous methods pay more attention to the design of visual feature extractor and feature fusion module, while employing U-Net \cite{Ronneberger15} as the default sound processing network. (b) Our approach inherently integrates feature fusion and sound signal processing into one network (i.e., PCNet), which alleviates the need for sophisticated visual guidance, while achieving competitive performance in a parameter efficient manner.}
	\label{fig:compare_framework}	
\end{figure}
Today's dominant paradigm for visual sound separation is constituted mainly by three components, i.e., visual feature extraction, multimodal feature fusion, and sound signal processing, as shown in Fig. \subref*{fig:compare_framework_previous}. The visual feature extraction part, drawing a lot of research attention and various model extensions exist, acts to derive discriminative visual features of the sounding object from video frames. Such discriminative features could be, for example, visual appearance cues extracted by an object recognition model \cite{Zhao18}, pixel-wise trajectories computed based on dense optical flow \cite{Zhao19,Zhu20a}, body and hand dynamics formulated with keypoint estimation \cite{Gan20}, object proposals obtained by a pre-trained object detector \cite{Gao19,Chatterjee21}, \emph{etc}. As for multimodal feature fusion, a simple and widely-used strategy is concatenating audio and visual features alone the channel dimension \cite{Gao19,Hu20,Chatterjee21,Chatterjee22}. Several works also exploit other fusion forms, such as the feature-wise affine transformation (FiLM) \cite{Perez18,Zhao19} and the self-attention based cross-modal fusion \cite{Gan20,Zhu20a}. In terms of sound signal processing, the U-Net \cite{Ronneberger15} style encoder-decoder network is usually treated as a default model in the majority of existing works \cite{Zhao18,Gao19,Xu19,Zhao19,Gan20,Zhu20a,Chatterjee21,Chatterjee22}.

Despite the success of the divide-and-conquer paradigm mentioned above, there are still some concerns left unaddressed. (i) The complete pipeline is parameter inefficient. While sophisticated visual guidance usually provides discriminative cues, it comes at the price of relying heavily on various well-trained vision models \cite{Zhao19,Gao19,Zhu20a,Chatterjee22} or third-party toolboxes \cite{Gan20}. As a result, the combination of those (deep neural network) modules makes the entire model cumbersome. For instance, to separate sounds with motion cues, Gan \emph{et al.} \cite{Gan20} design a pipeline consisting of seven modules: ResNet-50 for global semantic features, AlphaPose toolbox for human body keypoints, a hand detector and OpenPose for hand keypoints, a graph CNN to fuse semantic context and body dynamics, self-attention module for audio-visual fusion, and U-Net style architecture for sound processing. (ii) It is challengeable to optimize different model parts jointly and effectively. Since different pre-trained networks work with distinct dynamics and settings (e.g., optimizer, learning rate, weight decay, \emph{etc}), fine tuning the composition version of them becomes fragile, which may produce suboptimal sound separation results. (iii) It is open to question that U-Net is commonly viewed as a plausibly versatile sound analysis model for the visual sound separation task. The fused audio-visual features only serve as input to the U-Net decoder, and thus the interaction between fused features and low-level audio features (from U-Net encoder) implicitly occurs through the skip connection and transposed convolution. It is worth investigating whether there are other more effective ways to engage audio features with audio-visual information at different levels of abstraction.

We address above concerns by proposing a novel visually-guided sound source separation approach, named audio-visual predictive coding (AVPC). The AVPC model is constructed based on two networks: video analysis network and sound separation network, with no need to separately design feature fusion module. Specifically, as shown in Fig. \subref*{fig:compare_framework_ours}, object appearance features are first extracted from video frames by the simple video analysis network. Then, the sound separation network acts to estimate sound separation masks with the visual cues. The critical insight underlying our sound separation network is that different modalities' features of the same sounding object should share the same semantics in deep latent space, and hence should be predictable with each other. To this end, inspired by predictive coding (PC) in neuroscience \cite{Rao99,Friston05,Spratling17,Wen18,Song18}, the sound separation network predicts visual features with audio features extracted from mixture sound spectrogram, and improves the separation accuracy via iteratively minimizing the prediction error. More importantly, the PC-based sound separation network implements audio feature extraction, multimodal feature fusion, and sound separation mask prediction with the same architecture, which is more parameter efficient than previous paradigm. After that, the progressively refined separation masks are utilized to recover sound components of interest. We also devise a self-supervised audio-visual representation learning strategy for AVPC, called representation co-prediction (RCoP). The RCoP premises that if two mixture sounds include a same single sound component (e.g., a sound clip of guitar is mixed with the sound of trumpet and the sound of saxophone, respectively), the two groups of audio-visual representations corresponding to that sound component (e.g., guitar) should be similar. As a result, reducing the distance between the two groups of representations makes sound separation task easier.

In summary, our contributions are threefold:
\begin{enumerate}
	\item We propose a novel approach called AVPC for visually-guided sound source separation. Unlike previous arts investigating sophisticated visual feature extraction and multimodal feature fusion, AVPC focuses on sound signal processing with a bi-directional network architecture and iterative inference mechanism. This shift in perspective provides a parameter efficient and more effective way to handle visual sound separation task.
	
	\item We extend a self-supervised visual representation learning method to audio-visual setting, specifically to visual sound separation, which can be adopted as a pretext task to further boost sound separation performance.
	
	\item We systematically show that by adopting solely the plain visual appearance features, AVPC outperforms competitors that also use only appearance features as visual guidance, while achieves performance equivalent to those detection-based and motion-based methods.
\end{enumerate}

\section{Related Work}\label{sec:related_work}
\subsection{Sound Source Separation}
Sound source separation (from pure audio input), known as the ``cocktail party problem'' \cite{Haykin05} in the sound signal processing field, has been investigated extensively in the past few years. By assuming that the sound spectrogram has a low-rank structure, methods based on non-negative matrix factorization (NMF) \cite{Lee00} have historically been the most prominent on this task, as seen in \cite{Smaragdis03,Schmidt06,Virtanen07,Spiertz09,Cichocki09}. However, these methods as shallow models cannot deal with the high nonlinearity of sounds. By contrast, deep learning methods proposed recently give promising solutions to this problem \cite{Huang14,Hershey16,Chandna17,Yu17,Stoller18,Luo19}. Representative works for monaural speech separation include the model built on deep recurrent neural networks \cite{Huang14}, and the specifically-designed convolutional neural networks with vertical and horizontal convolution operations \cite{Chandna17}. To mitigate the long-standing label ambiguity problem, the deep clustering method \cite{Hershey16} and the permutation invariant training criterion \cite{Yu17} were designed, respectively, and both of which generalized well over unknown speakers and languages. A detailed discussion on this research direction can be found in \cite{Wang18}. Note that different from all of the above, we use the visual features extracted from available video frames to guide sound source separation.
 
\subsection{Visually-Guided Sound Source Separation}
With the aid of visual cues, additional discriminative features of target sound makers can be provided to the sound separation system, which proves to be beneficial to yield more accurate sound predictions \cite{Michelsanti21}. One line of work explored audio-visual speech separation, where the information of face recognition embeddings \cite{Ephrat18}, facial movements \cite{Gabbay18}, lip motions \cite{Afouras18,Lu18}, and face landmarks’ movements \cite{Morrone19} was employed as visual guidance, respectively.

More recently, another line of work focused on visually-guided sound source separation for non-speech signals. Two contemporaneous and pioneering works for this task come from Gao \emph{et al.} \cite{Gao18} and Zhao \emph{et al.} \cite{Zhao18}, who presented the multi-instance multi-label (MIML) and the sound-of-pixels (SoP) systems, respectively. The MIML \cite{Gao18} offered a learning framework to match audio bases with the object category predictions, where the learned audio bases proceed to supervise the NMF-based separation process. In SoP \cite{Zhao18}, an audio-visual two-stream network was proposed and trained by the mix-and-separate self-supervised learning procedure. The impressive performance of SoP intrigued the research community to make further efforts to address more challenging problems in this area. First, to tackle the homo-musical separation problem, where different musical instruments belonging to the same category emit sound simultaneously, motion-related visual cues were usually used to capture the dynamic difference in vision. These motion cues could be the trajectory of pixels \cite{Zhao19}, keypoint-based body dynamics \cite{Gan20}, optical flow \cite{Chatterjee22}, their combinations \cite{Zhu20a}, \emph{etc}. Second, to distinguish sound components with arbitrary numbers and types, Xu \emph{et al.} \cite{Xu19} proposed a novel two-stage system named minusplus network (MP-Net), where the minus stage recursively separates and removes the sound with highest energy, and the plus stage refines each separated sound accordingly. Third, to alleviate the unreality issue raised by training with artificially mixed video clips, the co-separation approach \cite{Gao19} was devised, which learns an association between consistent sounds and similar-looking objects across pairs of training videos.

Compared with aforementioned methods, our AVPC implements audio feature extraction, multimodal feature fusion, and sound separation mask prediction with the same PC-based architecture, which turns out to be more effective and parameter efficient over these U-Net-based ones \cite{Zhao18,Gao19,Xu19,Zhao19,Gan20,Zhu20a,Chatterjee21,Chatterjee22}. Additionally, the proposed RCoP, as a new task-oriented audio-visual representation learning strategy, extends the widely applied mix-and-separate training paradigm \cite{Zhao18}, leading to improved sound separation quality.

\subsection{Audio-Visual Learning}
There is increased interest in leveraging audio-visual correspondence to learn multimodal representations in a self-supervised manner. The widely-used correspondence is the semantic consistency across modalities \cite{Aytar16,Owens16,Arandjelovic17,Arandjelovic18}, i.e., audio and visual features extracted from the same video clip should have the same semantic category. In this regard, the feature of one modality with certain category could be treated as supervisory signal to guide representation learning on another modality \cite{Aytar16,Owens16}. Besides, the temporal synchronization between audio and visual streams is also useful \cite{Korbar18,Owens18}. As show in \cite{Korbar18}, audio and visual samples taken from different slices of the same video were viewed as hard negatives, and performing contrastive learning with such hard negatives resulted in powerful multi-sensory representations. What's more, some other works also explored the audio-visual correspondence from a variety of interesting perspectives, including feature clustering \cite{Hu19,Alwassel20,Hu20}, semantic comparisons among triple modalities in vector embedding space \cite{Alayrac20}, and audio-visual object embedding learning \cite{Afouras20}. A comprehensive survey on this topic can be found in \cite{Zhu21}.

The representations learned by these methods are task-agnostic, which provide an effective initialization to improve performance of various downstream tasks, such as action/scene recognition \cite{Owens16,Korbar18,Owens18,Arandjelovic17,Alwassel20,Alayrac20}, temporal action localization \cite{Alwassel20}, audio classification \cite{Aytar16,Korbar18,Arandjelovic17,Hu19,Alwassel20,Alayrac20}, sound source localization \cite{Owens18,Arandjelovic18,Hu19,Hu20,Afouras20} and separation \cite{Owens18,Hu20,Afouras20}, \emph{etc}. By contrast, our proposed RCoP, as a task-oriented self-supervised learning strategy, serves to learn representations customized for sound source separation, and thus is more effective. In addition, by taking inspiration from the self-supervised visual representation learning paradigms, BYOL \cite{Grill20} and SimSiam \cite{Chen21c}, the whole network can be trained by RCoP without using negative samples, large batches, and momentum encoders. Therefore, RCoP offers an economical way to learn audio-visual representations from unlabeled videos.

\section{Preliminaries}\label{sec:pcnet_intro}
In this section, we give a concise introduction to the predictive coding network (PCNet) \cite{Wen18} that inspires us to construct sound separation network. We embody the information processing mechanism in PCNet by formulating its optimization objective and representation updating rules, respectively.

\subsection{Core Idea and Optimization Objective of PCNet}
\begin{figure}
	\centering
	\includegraphics[width=0.9\linewidth]{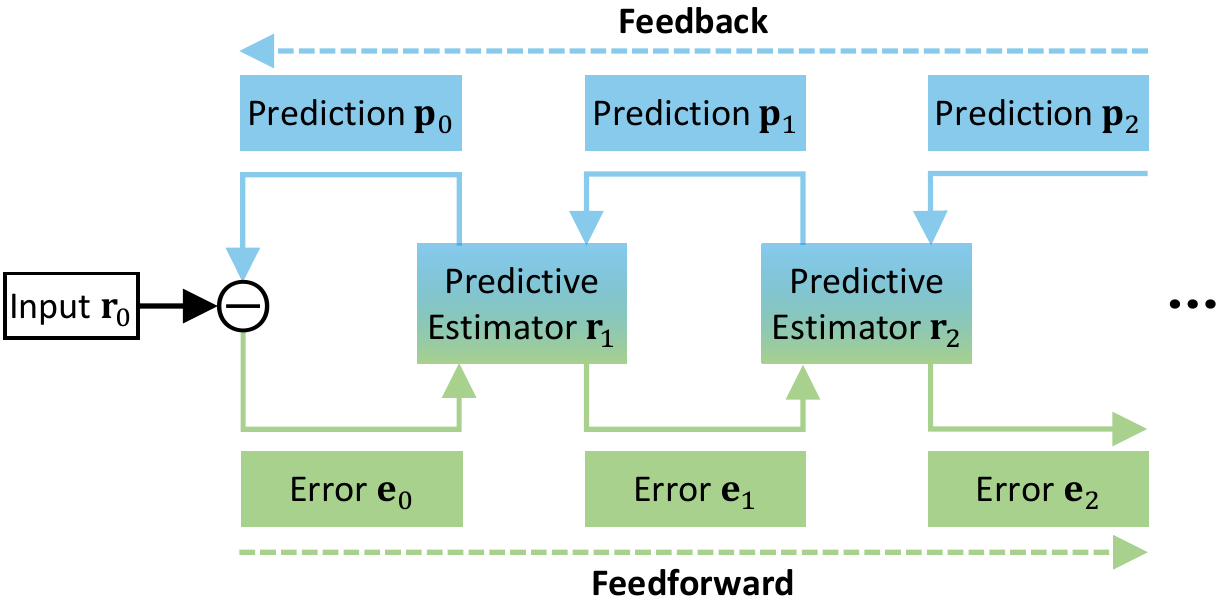}
	\caption{General architecture of the hierarchical predictive coding framework (recreated based on the PC sketch in \cite{Rao99}). At each hierarchical level, the feedback connections carry predictions of neural activities at the lower level, whereas feedforward connections carry prediction errors between the predictions and the actual lower-level activities. The predictive estimator uses both the prediction errors from lower level and the predictions from higher level to correct its hidden representations.}
	\label{fig:pc_sketch}
\end{figure}
In fact, at the heart of PCNet is the predictive coding (PC) model in neuroscience \cite{Rao99,Friston05,Spratling17,Wen18,Song18}. As a specific neural network model (see Fig. \ref{fig:pc_sketch}), PC uses feedback connections from a higher area to a lower area (e.g., V2 to V1 in visual cortex) to convey predictions of lower-level neural activities; while employs feedforward connections to carry prediction errors between the predictions and the actual lower-level activities; and the feedback and feedforward computing processes are executed alternatively such that prediction errors are dynamically reduced and all layers' representations are progressively refined. The basic PC in \cite{Rao99} performs alternative computing in local level, i.e., the representation of current layer starts to be updated only when the previous layer finishes the two computing processes. As one of the instantiations of PC, PCNet disentangles the two processes in global level, meaning that representations of all layers are first modulated in feedback (or feedforward) process, and then updated again in feedforward (or feedback) counterpart. By doing so, PCNet achieves competitive and more definitive object recognition results with less parameters compared with convolutional feedforward-only neural networks \cite{Wen18}.

Formally, for the $l$-th hidden layer, we denote the representation as $\mathbf{r}_{l}$, and the feedback connection weights from layer $l$ to layer $l-1$ as $\mathbf{W}_{l,l-1}$ (similarly $\mathbf{W}_{l-1,l}$). At layer $l$, the optimization objective is to minimize the compound loss:
\begin{equation}
	\mathcal{L}^{l} = \frac{\alpha_{l}}{2}\underbrace{\|\mathbf{r}_{l-1}-g((\mathbf{W}_{l,l-1})^{\mathrm{T}}\mathbf{r}_{l})\|^{2}_{2}}_{\mathcal{L}_{1}^{l}} + \frac{\beta_{l}}{2}\underbrace{\|\mathbf{r}_{l} - \mathbf{p}_{l}\|^{2}_{2}}_{\mathcal{L}_{2}^{l}},\label{eqn:loss_pcnet}
\end{equation}
where the function $g$ stands for a generative transformation, $\alpha_{l}$ and $\beta_{l}$ are hyperparameters controlling the relative importance of the two loss terms $\mathcal{L}_{1}^{l}$ and $\mathcal{L}_{2}^{l}$, and $\mathbf{p}_{l}=g((\mathbf{W}_{l+1,l})^{\mathrm{T}}\mathbf{r}_{l+1})$ is the prediction of $\mathbf{r}_{l}$. Here, the prediction $\mathbf{p}_{l}$ is derived through a nonlinear transformation on the higher layer representation $\mathbf{r}_{l+1}$, and can be viewed as prior knowledge based on past learning experiences \cite{Spratling17}. From \eqref{eqn:loss_pcnet} we see that the optimal representation of layer $l$, $\mathbf{r}_{l}$, can not only reconstruct the previous layer representation $\mathbf{r}_{l-1}$ (by minimizing $\mathcal{L}_{1}^{l}$), but also be similar with the prediction $\mathbf{p}_{l}$ from higher layer (by minimizing $\mathcal{L}_{2}^{l}$). Consequently, each layer representation is mediated so as to adaptively fuse input information (from lower-level $\mathbf{r}_{l-1}$) and prior knowledge (from higher-level $\mathbf{p}_{l}$).

\subsection{Basic Rules of Representation Updating in PCNet}
As for representation updating in \emph{feedback} process, PCNet uses the prediction signal generated from higher layer to adjust the representation of current layer. Similar to \cite{Rao99,Wen18}, we set $g$ as a liner-generative transformation to simplify derivation, i.e., $g((\mathbf{W}_{l,l-1})^{\mathrm{T}}\mathbf{r}_{l})=(\mathbf{W}_{l,l-1})^{\mathrm{T}}\mathbf{r}_{l}$. Then the prediction signal of $\mathbf{r}_{l}$ at time step $t$ is computed as:
\begin{equation}
	\mathbf{p}_{l}(t)=(\mathbf{W}_{l+1,l})^{\mathrm{T}}\mathbf{r}_{l+1}(t).\label{eqn:pred_linear}
\end{equation}
After that, we substitute \eqref{eqn:pred_linear} into the loss term $\mathcal{L}_{2}^{l}$ in \eqref{eqn:loss_pcnet}, and minimize $\mathcal{L}_{2}^{l}$ w.r.t. $\mathbf{r}_{l}$ by gradient decent, leading to following update rules:
\begin{align}
	\frac{\partial \mathcal{L}_{2}^{l}}{\partial \mathbf{r}_{l}(t)} &= 2(\mathbf{r}_{l}(t) - \mathbf{p}_{l}(t)), \\
	\mathbf{r}_{l}(t+1) &= \mathbf{r}_{l}(t) - \eta_{l}\frac{\beta_{l}}{2}\frac{\partial \mathcal{L}_{2}^{l}}{\partial \mathbf{r}_{l}(t)} \nonumber\\
	&= (1 - \eta_{l}\beta_{l})\mathbf{r}_{l}(t) + \eta_{l}\beta_{l}\mathbf{p}_{l}(t), \label{eqn:pcnet_feedback_repres0}
\end{align}
where the non-negative scalar $\eta_{l}$ governs feedback representation learning. For simplicity, we set $b_{l}=\eta_{l}\beta_{l}$ and rewrite \eqref{eqn:pcnet_feedback_repres0} as follows:
\begin{equation}
	\mathbf{r}_{l}(t+1) = (1 - b_{l}) \mathbf{r}_{l}(t) + b_{l} \mathbf{p}_{l}(t).\label{eqn:pcnet_feedback_repres1}
\end{equation}

On the other hand, in the \emph{feedforward} process PCNet updates the representation again based on the prediction error conveyed from previous layer. The prediction error of layer $l-1$ is defined as:
\begin{equation}
	\mathbf{e}_{l-1}(t) = \mathbf{r}_{l-1}(t) - \mathbf{p}_{l-1}(t), \label{eqn:pred_error}
\end{equation}
which manifests the difference between the actual representation and its prediction. By using gradient decent to minimize $\mathcal{L}_{1}^{l}$ w.r.t. $\mathbf{r}_{l}$, we derive the update equation:
\begin{align}
	\frac{\partial \mathcal{L}_{1}^{l}}{\partial \mathbf{r}_{l}(t)} &= -2\mathbf{W}_{l,l-1}\mathbf{e}_{l-1}(t), \\
	\mathbf{r}_{l}(t+1) &= \mathbf{r}_{l}(t) - \kappa_{l}\frac{\alpha_{l}}{2}\frac{\partial \mathcal{L}_{1}^{l}}{\partial \mathbf{r}_{l}(t)} \nonumber\\
	&= \mathbf{r}_{l}(t) + \kappa_{l}\alpha_{l}\mathbf{W}_{l,l-1}\mathbf{e}_{l-1}(t), \label{eqn:pcnet_feedforward_repres0}
\end{align}
where $\kappa_{l}$ is a scalar like $\eta_{l}$. We also set $a_{l}=\kappa_{l}\alpha_{l}$ to simplify denotation. Following \cite{Wen18}, we introduce more degrees of freedom to representation learning by assuming that feedback connection weights are the transposed feedward connection weights, i.e., $\mathbf{W}_{l,l-1}=(\mathbf{W}_{l-1,l})^{\mathrm{T}}$. As a result, the update equation in \eqref{eqn:pcnet_feedforward_repres0} is equivalently converted to the following form:
\begin{equation}
	\mathbf{r}_{l}(t+1) = \mathbf{r}_{l}(t) + a_{l} (\mathbf{W}_{l-1,l})^{\mathrm{T}}\mathbf{e}_{l-1}(t).\label{eqn:pcnet_feedforward_repres1}
\end{equation}

To add nonlinearity to the above two process, we can apply some nonlinear activation function (e.g., \texttt{ReLU} \cite{Nair10} as used in \cite{Wen18}) to the output of each convolutional layer. Therefore, the final nonlinear feedback process is described as:
\begin{equation}
	\mathbf{r}_{l}(t+1) = \texttt{ReLU}((1 - b_{l}) \mathbf{r}_{l}(t) + b_{l} \mathbf{p}_{l}(t)),\label{eqn:pcnet_feedback_repres2}
\end{equation}
and the nonlinear feedforward process is given as:
\begin{equation}
	\mathbf{r}_{l}(t+1) = \texttt{ReLU}(\mathbf{r}_{l}(t) + a_{l} (\mathbf{W}_{l-1,l})^{\mathrm{T}}\mathbf{e}_{l-1}(t)).\label{eqn:pcnet_feedforward_repres2}
\end{equation}

PCNet performs these two processes alternatively several times, and consequently all layers' representations are refined in a recursive manner. In Section \ref{sec:sound_separation_network}, we will extend the basic representation updating rules to drive learning in predictive coding-based sound separation network.

\section{Proposed Method}\label{sec:proposed_method}
\begin{figure*}
	\centering
	\includegraphics[width=\linewidth]{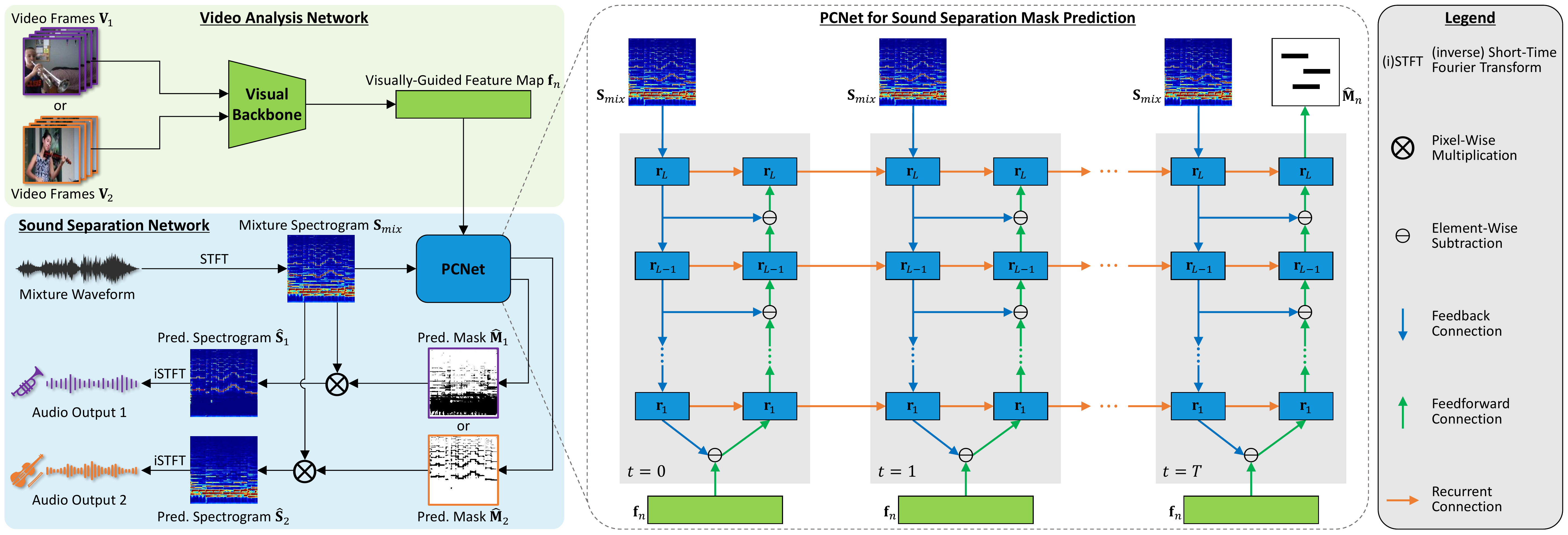}
	\caption{Architecture of the audio-visual predictive coding (AVPC). Our framework consists of two branches. The video analysis network branch focuses on extracting semantic visual features from video frames. The sound separation network branch takes as input mixture spectrogram and visually-guided feature map, and is responsible to predict sound separation mask. Here we linearly sum single sound source waveforms from different video clips to simulate mixture waveform, which is then converted to mixture spectrogram by STFT. The sound separation network features the PCNet that fuses multimodal information in an iterative fashion. The whole network is trained by minimizing the per-pixel binary cross-entropy between predicted masks and ground truth masks.}\label{fig:avpc_archi}
\end{figure*}

\subsection{Overview}
The key insights underlying our method are twofold. On the one hand, by leveraging audio feature to predict semantic visual feature, AVPC implements audio feature extraction, multimodal feature fusion, and sound separation mask prediction in the same one network architecture (i.e., the sound separation network). The fusion results can be increasingly refined with the inherent representation updating rules, i.e., iteratively minimizing the prediction error in bi-directional processes. Equipped with such network architecture and updating rules, we mitigate the need for laborious design for and fragile optimization on various model components. On the other hand, the self-supervised learning strategy RCoP works to excavate the correlations between two sound mixtures that include a same sound source. To this end, RCoP makes the two learned audio-visual representations corresponding to the shared sound source as similar as possible.

From the view of model structure, AVPC consists of two network branches, i.e., video analysis network and sound separation network, as shown in Fig. \ref{fig:avpc_archi}. First, the video analysis network takes as input video frames $\mathbf{V}_{n}$ and acts to compute semantic feature map as visually-guided signal $\mathbf{f}_{n}$\footnote{Hereafter we use $n\in\{1,2,\dots,N\}$ to index different video clips.}. Then, the sound separation network extracts audio features from the mixture sound spectrogram $\mathrm{\mathbf{S}}_{mix}$, and uses these extracted features to predict the visually-guided feature map recursively, resulting in the progressively refined audio-visual representation $\mathbf{r}_{L}(T)$. Next, a transposed convolutional layer followed by a sigmoid function (omitted in Fig. \ref{fig:avpc_archi} for simplicity) is operated on the audio-visual representation, obtaining a prediction of the sound separation mask $\hat{\mathbf{M}}_{n}$. Finally, performing the pixel-wise multiplication between the predicted mask and the mixture spectrogram outputs a specific sound spectrogram $\hat{\mathrm{\mathbf{S}}}_{n}$, which is further converted to sound signal by applying an inverse Short-Time Fourier Transform (iSTFT).

In the following, we detail the two network branches, training methods, and representation co-prediction strategy of AVPC, respectively.

\subsection{Video Analysis Network}
The deep features of video frames are extracted by the video analysis network which, similar to a series of existing works \cite{Zhao18,Gao19,Xu19,Zhao19,Zhu20a,Qian20}, takes the plain ResNet-18 \cite{He16} as a backbone. Given a video clip of size $\mathrm{F}\times\mathrm{H}\times\mathrm{W}\times3$, where $\mathrm{F}$, $\mathrm{H}$, and $\mathrm{W}$ denote number, height, and width of video frames, respectively, the ResNet extracts high-level semantic features from each frame before the spatial average pooling layer, obtaining a feature tensor of size $\mathrm{F}\times(\mathrm{H}/32)\times(\mathrm{W}/32)\times512$. Subsequently, by conducting an additional convolution operation spatially and then temporal pooling along the first time dimension $\mathrm{F}$, the feature tensor will be transformed into a smaller version of size $\hat{\mathrm{H}}\times\hat{\mathrm{W}}\times\mathrm{K}$, where $\hat{\mathrm{H}}$, $\hat{\mathrm{W}}$, and $\mathrm{K}$ stand for height, width, and number of channels of the final visually-guided feature map $\mathbf{f}_{n}$, respectively. In our all implementations, we set $\hat{\mathrm{H}}=2$, $\hat{\mathrm{W}}=2$, and $\mathrm{K}=16$.

We empirically show that although AVPC is conditioned on the visually-guided feature map that only captures sound makers' appearance information, it can still perform on par with detection-based \cite{Gao19} and motion-based \cite{Zhao19} visual sound separation methods (see quantitative results in Section \ref{sec:quantitative_exp}).

\subsection{Predictive Coding-Based Sound Separation Network}\label{sec:sound_separation_network}
\subsubsection{Formulation}
We extend PCNet introduced in Section \ref{sec:pcnet_intro} to construct the sound separation network, which is mainly characterized by a cross-modal feature prediction mechanism. The assumption supporting this prediction mechanism is that audio and visual information extracted from the same sounding object should be consistent in high-level semantic feature space, and thus could be predictable with each other. In this regard, we develop the iterative minimization process of PCNet to gradually decrease the prediction error between audio and visual features.

Formally, sound separation network receives the visually-guided feature map $\mathbf{f}_{n}$ and the mixture spectrogram $\mathrm{\mathbf{S}}_{mix}$ as input, where $\mathbf{f}_{n}$ is viewed as the target signal to be predicted, and $\mathrm{\mathbf{S}}_{mix}$ is derived from the mixture waveform by applying a Short-Time Fourier Transform (STFT). Let $L$ denote the number of layers, and $T$ for the recursive cycles. The sound separation network uses the following steps to recursively update representations (i.e., fuse audio and visual features).

First, it initializes representations of all layers, $\{\mathbf{r}_{l}^{\text{init}}|l=1,2,\cdots,L\}$, in a \emph{top-down} manner (from layer $L$ to layer 1):
\begin{equation}\label{eqn:init_r_fb_avpc}
	\mathbf{r}_{l}^{\text{init}} = g((\mathbf{W}_{l+1,l})^{\mathrm{T}}\mathbf{r}_{l+1}^{\text{init}}),
\end{equation}
where we set $\mathbf{r}_{L+1}^{\text{init}}\triangleq\mathrm{\mathbf{S}}_{mix}$ and instantiate the generative transformation $g$ with nonlinear function \texttt{LeakyReLU} \cite{Maas13}. Starting at the mixture spectrogram, this top-down initialization actually extracts audio features layer-by-layer, and thus injects sound-related information into each layer's representation, from a local spatial scale (i.e., $\mathbf{r}_{L}^{\text{init}}$) to a global spatial scale (i.e., $\mathbf{r}_{1}^{\text{init}}$).

Then, we are able to update all representations based on $\{\mathbf{r}_{l}^{\text{init}}|l=1,2,\cdots,L\}$ and $\mathbf{f}_{n}$ for time $t=0$. To this end, a \emph{bottom-up} error propagation procedure, similar to the nonlinear feedforward process of PCNet in \eqref{eqn:pcnet_feedforward_repres2}, is triggered and formulated as follows (from layer 1 to layer $L$):
\begin{align}\label{eqn:init_r_ff_avpc}
	\mathbf{e}_{l-1}(0) &= \mathbf{r}_{l-1}(0) - \mathbf{r}_{l-1}^{\text{init}},\\
	\mathbf{r}_{l}(0) &= f(\mathbf{r}_{l}^{\text{init}} + a_{l} (\mathbf{W}_{l-1,l})^{\mathrm{T}}\mathbf{e}_{l-1}(0)),
\end{align}
where $\mathbf{r}_{0}(t)\equiv\mathbf{f}_{n}$, $\mathbf{r}_{0}^{\text{init}}=g((\mathbf{W}_{1,0})^{\mathrm{T}}\mathbf{r}_{1}^{\text{init}})$ is the coarse prediction of $\mathbf{f}_{n}$, and $f$ is a nonlinear activation function (e.g., \texttt{LeakyReLU}) in the feedforward process.

Up to now, sound separation network finishes one recursive cycle to update all representations, where the top-down feedback process only works for representation initialization (i.e., \eqref{eqn:init_r_fb_avpc}). To proceed the update rules in both top-down feedback and bottom-up feedforward processes alternatively, we perform the following calculations.

\noindent \emph{Nonlinear feedback process ($l=L, L-1, \cdots, 1$):}
\begin{align}
	\mathbf{p}_{l}(t) &=(\mathbf{W}_{l+1,l})^{\mathrm{T}}\mathbf{r}_{l+1}(t), \label{eqn:update_p_fb_avpc}\\
	\mathbf{r}_{l}(t+1) &= g((1 - b_{l}) \mathbf{r}_{l}(t) + b_{l} \mathbf{p}_{l}(t)), \label{eqn:update_r_fb_avpc}
\end{align}
where $\mathbf{r}_{L+1}(t)\equiv\mathrm{\mathbf{S}}_{mix}$.

\noindent \emph{Nonlinear feedforward process ($l=1, 2, \cdots, L$):}
\begin{align}
\mathbf{e}_{l-1}(t) &= \mathbf{r}_{l-1}(t) -  \mathbf{p}_{l-1}(t), \label{eqn:update_e_ff_avpc}\\
\mathbf{r}_{l}(t+1) &= f(\mathbf{r}_{l}(t) + a_{l} (\mathbf{W}_{l-1,l})^{\mathrm{T}}\mathbf{e}_{l-1}(t)), \label{eqn:update_r_ff_avpc}
\end{align}
where $\mathbf{p}_{0}(t)=g((\mathbf{W}_{1,0})^{\mathrm{T}}\mathbf{r}_{1}(t))$.

\begin{algorithm}[t]
	\caption{Mask Prediction with Predictive Coding}\label{alg:recursive_comput}
	\begin{algorithmic}[1]
		\REQUIRE mixture sound spectrogram $\mathrm{\mathbf{S}}_{mix}$; visual feature $\mathbf{f}_{n}$
		\ENSURE sound separation mask $\hat{\mathrm{\mathbf{M}}}_{n}$
		\STATE $\mathbf{r}_{L+1}^{\text{init}} \leftarrow \texttt{BN}(\mathrm{\mathbf{S}}_{mix})$
		\STATE $\mathbf{r}_{L+1}(t) \leftarrow \texttt{BN}(\mathrm{\mathbf{S}}_{mix})$
		\STATE $\mathbf{r}_{0}(t) \leftarrow \mathbf{f}_{n}$
		\STATE \COMMENT{initialize representations}
		\FOR {$l = L$ to 1}
		\STATE $\mathbf{r}_{l}^{\text{init}} \leftarrow \texttt{LeakyReLU}(\texttt{BN}(\texttt{Conv}(\mathbf{r}_{l+1}^{\text{init}})))$
		\ENDFOR
		\STATE $\mathbf{r}_{0}^{\text{init}} \leftarrow \texttt{LeakyReLU}(\texttt{Conv}(\mathbf{r}_{1}^{\text{init}}))$			
		\FOR {$l=1$ to $L$}
		\STATE $\mathbf{e}_{l-1}(0) \leftarrow \mathbf{r}_{l-1}(0) - \mathbf{r}_{l-1}^{\text{init}}$
		\STATE {\small $\mathbf{r}_{l}(0) \leftarrow \texttt{LeakyReLU}(\texttt{BN}(\mathbf{r}_{l}^{\text{init}} + a\texttt{TransConv}(\mathbf{e}_{l-1}(0))))$}
		\ENDFOR
		\STATE \COMMENT{recurrent computation with $T$ cycles}
		\FOR {$t = 1$ to $T$}
		\STATE \COMMENT{nonlinear feedback process}
		\FOR {$l = L$ to 1}
		\STATE $\mathbf{p}_{l}(t-1) \leftarrow \texttt{Conv}(\mathbf{r}_{l+1}(t-1))$
		\STATE {\footnotesize $\mathbf{r}_{l}(t-1) \leftarrow \texttt{LeakyReLU}(\texttt{BN}((1 - b)\mathbf{r}_{l}(t-1) + b\mathbf{p}_{l}(t-1)))$}
		\ENDFOR
		\STATE $\mathbf{p}_{0}(t-1) \leftarrow \texttt{LeakyReLU}(\texttt{Conv}(\mathbf{r}_{1}(t-1)))$
		\STATE \COMMENT{nonlinear feedforward process}
		\FOR {$l = 1$ to $L$}
		\STATE $\mathbf{e}_{l-1}(t) \leftarrow \mathbf{r}_{l-1}(t) -  \mathbf{p}_{l-1}(t-1)$
		\STATE {\footnotesize $\mathbf{r}_{l}(t) \leftarrow \texttt{LeakyReLU}(\texttt{BN}(\mathbf{r}_{l}(t-1) + a\texttt{TransConv}(\mathbf{e}_{l-1}(t))))$}
		\ENDFOR
		\ENDFOR
		\STATE \COMMENT{mask prediction}
		\STATE $\hat{\mathrm{\mathbf{M}}}_{n} \leftarrow \texttt{Sigmoid}(\texttt{TransConv}(\mathbf{r}_{L}(T)))$%
	\end{algorithmic}%
	\vspace{-0.8\baselineskip}
	\hrulefill \\
	\texttt{BN}: batch normalization; \texttt{Conv}: convolution; \texttt{TransConv}: transposed convolution; $a$ and $b$: specific and learnable parameters to each filter in each layer.
\end{algorithm}

The top-layer feature, $\mathbf{r}_{L}(t)$, can act as the audio-visual representation to predict sound separation mask. Due to the nature of iterative inference method adopted here, $\mathbf{r}_{L}(t)$ will evolve into a sequence of progressively refined audio-visual representations, i.e., $\{\mathbf{r}_{L}(t)|t=0,1,\cdots,T\}$. The output at last time step, $\mathbf{r}_{L}(T)$, is used to compute mask prediction $\hat{\mathbf{M}}_{n}$ with best performance. We summarize the whole process of sound separation mask prediction in Algorithm \ref{alg:recursive_comput}.

Finally, by multiplying the mixture spectrogram $\mathrm{\mathbf{S}}_{mix}$ with the predicted mask $\hat{\mathbf{M}}_{n}$, we obtain the $n$-th sound component spectrogram $\hat{\mathrm{\mathbf{S}}}_{n}$:
\begin{equation}\label{eqn:compute_sound_spect}
	\hat{\mathrm{\mathbf{S}}}_{n} = \mathrm{\mathbf{S}}_{mix} \otimes \hat{\mathbf{M}}_{n},
\end{equation}
where $\otimes$ denotes pixel-wise multiplication. The $n$-th sound component is then recovered through iSTFT applied on $\hat{\mathrm{\mathbf{S}}}_{n}$.

\subsubsection{Discussion}
The multimodal feature fusion in PCNet is reached by the iterative inference procedure. In fact, as formulated in Section \ref{sec:pcnet_intro}, updating representations based on \eqref{eqn:update_r_fb_avpc} and \eqref{eqn:update_r_ff_avpc} (or \eqref{eqn:pcnet_feedback_repres2} and \eqref{eqn:pcnet_feedforward_repres2}) is equivalent to minimizing prediction error of each layer. Particularly, reducing the bottom-layer prediction error, $\mathbf{e}_{0}(t)$, makes the prediction signal extracted from audio source, $\mathbf{p}_{0}(t)$, similar to the specific visual feature, $\mathbf{f}_{n}$, in the sense of $L_{2}$ distance. Because the visual feature carries discriminative information of sounding object, the prediction signal would ideally tend to be discriminative as well. In this regard, we think that the prediction from audio source is fused with visual information. Note that PCNet generates the prediction and propagates the prediction error layer-by-layer, therefore the semantic discrimination flows across hierarchy and the top-layer audio-visual representation, $\mathbf{r}_{L}(t)$, would become more discriminative as time goes on. In Section \ref{sec:ablation_iterative_infer}, we will empirically verify the effectiveness of this feature fusion mechanism by visualizing the semantic discrimination of learned embeddings.

\begin{figure}[t]
	\centering
	\subfloat[\label{fig:comp_u-net_pcnet_part1}]{\includegraphics[width=0.9\linewidth]{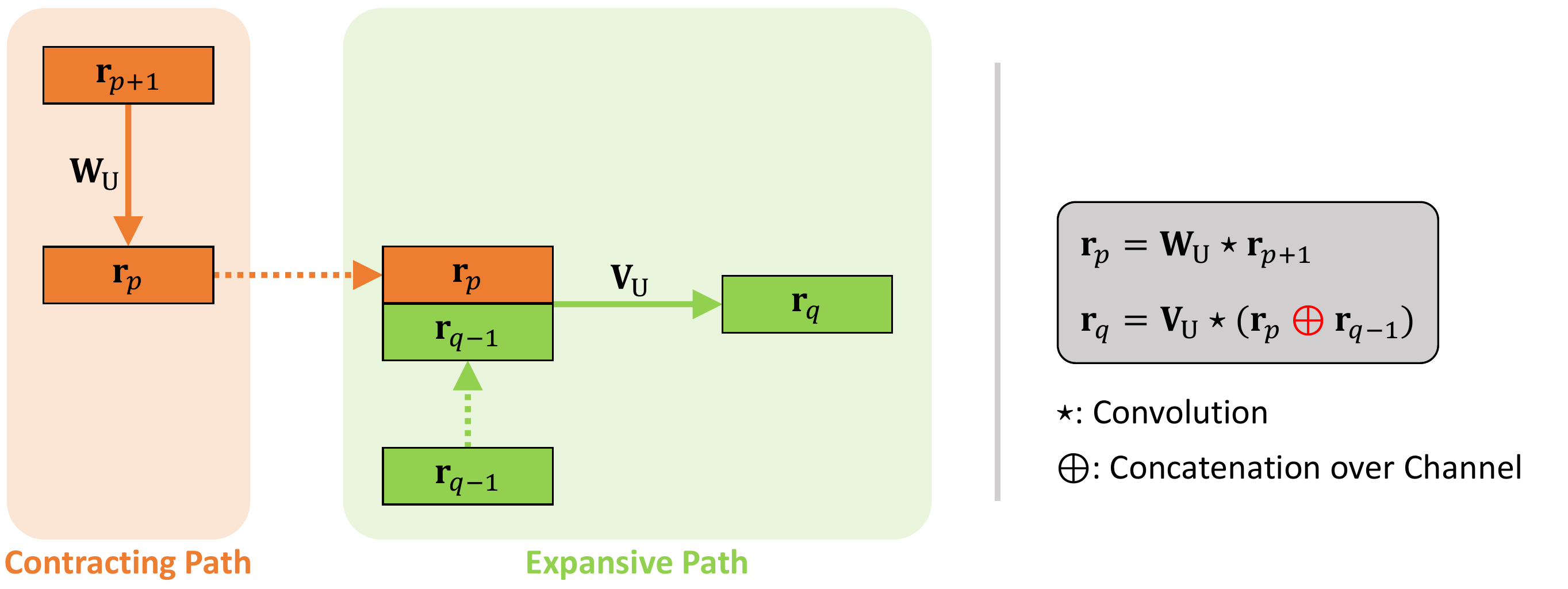}}\\
	\subfloat[\label{fig:comp_u-net_pcnet_part2}]{\includegraphics[width=0.9\linewidth]{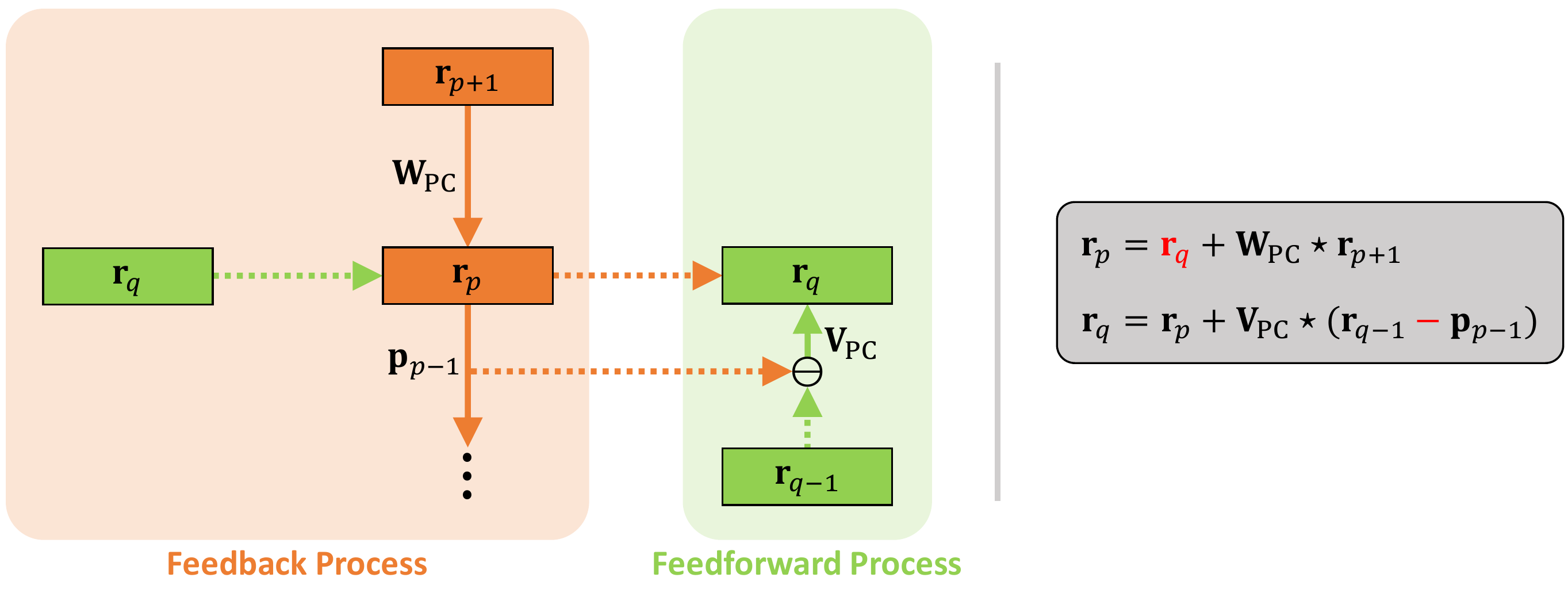}}
	\caption{Two ways to integrate different levels of features. (a) Basic module in U-Net. (b) Basic module in PCNet. In each panel, the subscripts $p$ and $q$ are indexes of two symmetric layers, solid lines denote weight connections between two layers, and dotted lines indicate duplication of features. For simplicity, we omit maxpooling and upsampling layers and meanwhile use linear activation function in both cases. Note that the concatenation $\oplus$ of $r_{p}$ and $r_{q-1}$ in (a) leads to enlarged feature map channel, and therefore requiring U-Net to use more filters to do convolution.}
	\label{fig:compar_u-net_pcnet}
\end{figure}
By comparing the basic modules constructing PCNet and U-Net, we find that the feedback and feedforward processes of PCNet in Fig. \subref*{fig:comp_u-net_pcnet_part2} are analogous to the contracting and expansive paths of U-Net in Fig. \subref*{fig:comp_u-net_pcnet_part1}, respectively. Equipped with this symmetric structure, both U-Net and PCNet provide ways to integrate different levels of features. For U-Net in Fig. \subref*{fig:comp_u-net_pcnet_part1}, the lower-level feature map $\mathbf{r}_{p}$ are concatenated with the higher-level one $\mathbf{r}_{q-1}$ over channel, and then converted to new feature $\mathbf{r}_{q}$ by convolution operation. By doing so, the context information from contracting path are combined with the upsampled output from expansive path, yielding precise pixel-wise prediction. However, the channel expansion also induces more weight connections to deal with the appended feature map. By contrast, PCNet in Fig. \subref*{fig:comp_u-net_pcnet_part2} fuses features from two processes in a recurrent way, the realization of which needs no parameterized connection between symmetric layers (instead resorting to the recurrent connection), and turns out to be more parameter efficient.

Note that the extended PCNet differs from the raw version \cite{Wen18} (denoted as PCN here) in the following respects. (i) We customize the input configuration of the network such that bottom layer takes as input visual feature map, while top layer deals with sound spectrogram as a kind of prior knowledge. By contrast, PCN only receives image as input at bottom layer. (ii) The distinction between input configurations induces that PCNet is inherently able to handle multimodal learning task while PCN is fit for unimodal image classification task. (iii) We technically introduce and adjust the batch normalization variant originally designed for recurrent neural networks \cite{Laurent16}, which proves to be conducive to stabilizing the multimodal learning process in PCNet. (iv) We empirically show that through cross-modal feature prediction in PCNet, the semantic discrimination can be conveyed from visual feature to audio-visual representation (cf., Fig. \ref{fig:tsne_fusion_w_time_step}). This intriguing property is not explored before in PCN and other related works.

\subsection{Training Method}
We employ the Mix-and-Separate (MaS) framework proposed in \cite{Zhao18} to train models. Because of the additivity of sound signals \cite{Zhao18}, mixture audio ground truth can be artificially created by linearly mixing sound signals from different video clips. The goal of our model is then to recover target sound component from mixture audio conditioned on the corresponding visual cues.

Formally, we randomly sample $N$ video clips $\{\mathbf{V}_{n}, \mathbf{S}_{n}\}_{n=1}^{N}$ from training dataset, where $\mathbf{V}_{n}$ and $\mathbf{S}_{n}$ denote video frames and sound signal of the $n$-th video clip, respectively. Subsequently we linearly combine these sound signals to synthesize sound mixture as $\mathbf{S}_{mix}=\frac{1}{N}\sum_{n=1}^{N}\mathbf{S}_{n}$. The video analysis network takes as input video frames $\mathbf{V}_{n}$ and is responsible to compute visual feature map $\mathbf{f}_{n}$, which proceeds to help the sound separation network estimate sound component $\hat{\mathrm{\mathbf{S}}}_{n}$ from $\mathbf{S}_{mix}$. In practice, the direct output of the sound separation network is a binary mask $\hat{\mathbf{M}}_{n}$, while the ground truth mask is determined based on whether the $n$-th sound component is the dominant component in the input mixed sound, i.e.,
\begin{equation}\label{eqn:generate_gt_mask}
	\mathbf{M}_{n}(u,v) = [\![\mathbf{S}_{n}(u,v) \geq \mathbf{S}_{mix}(u,v)]\!],
\end{equation}
where $(u,v)$ indicates the time-frequency coordinates in the sound spetrogram, and $[\![\texttt{criteria}]\!]$ denotes an indicator function whose value is 1 when the \texttt{criteria} is satisfied, otherwise 0. Finally, the model is trained by minimizing per-pixel binary cross-entropy (\texttt{BCE}) between the predicted masks and the ground truth masks: 
\begin{equation}\label{eqn:bce_loss}
	\mathcal{L}_{MaS} = \frac{1}{N}\sum_{n=1}^{N}\texttt{BCE}(\hat{\mathbf{M}}_{n}, \mathbf{M}_{n}).
\end{equation}

\subsection{Boosting Performance by Representation Co-Prediction}
\begin{figure}
	\centering
	\includegraphics[width=\linewidth]{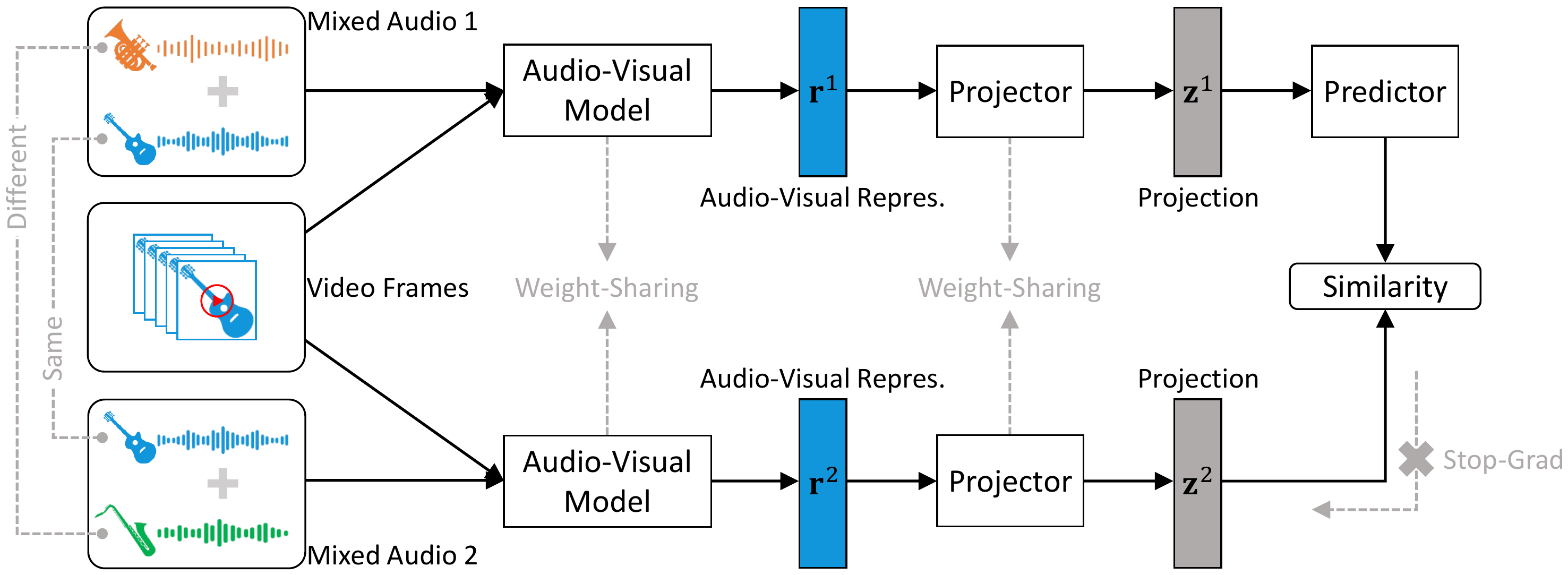}
	\caption{RCoP pipeline. The video frames and two mixed audios are processed by the same audio-visual model, respectively. Then a projector based on a multi-layer perceptron (MLP) transforms two audio-visual representations into their projection counterparts. A prediction MLP is further applied on one branch, and a stop-gradient operation is applied on the other branch. The whole model maximizes the similarity between both branches.}
	\label{fig:rcop}
\end{figure}
Inspired by the self-supervised visual representation learning methods BYOL \cite{Grill20} and SimSiam \cite{Chen21c}, we propose a new audio-visual representation learning strategy, named representation co-prediction (RCoP), to further boost the visual sound separation performance, as shown in Fig. \ref{fig:rcop}. Different from SimSiam that focuses on discovering semantic similarity between augmented views of the same image, the core idea of RCoP is to make two audio-visual representations of the given video clip (e.g., guitar playing here) as similar as possible. By doing that, RCoP enforces the audio-visual model to explore intrinsic correlation between two mixed audios sharing the same one sound component, and as a result providing a better parameter initialization for separating sound of interest.

Formally, given an video clip $(\mathbf{V}, \mathbf{S})$, we randomly select two audio clips from another two different videos, $\mathbf{S}^{1}$ and $\mathbf{S}^{2}$, and synthesize mixed audios $\mathbf{S}_{mix}^{1}=(\mathbf{S} + \mathbf{S}^{1}) / 2$ and $\mathbf{S}_{mix}^{2}=(\mathbf{S} + \mathbf{S}^{2}) / 2$, respectively. The new video pairs $(\mathbf{V}, \mathbf{S}_{mix}^{1})$ and $(\mathbf{V}, \mathbf{S}_{mix}^{2})$ are then fed into the audio-visual model like AVPC, obtaining two audio-visual representations $\mathbf{r}^{1}$ and $\mathbf{r}^{2}$, respectively. A projector further transforms the two representations into new latent space, producing projections $\mathbf{z}^{1}$ and $\mathbf{z}^{2}$, respectively. And a predictor head \texttt{Pred} takes as input $\mathbf{z}^{1}$ to approximate $\mathbf{z}^{2}$ by minimizing the following negative cosine similarity (NCS): 
\begin{equation}\label{eqn:loss_neg_cos_sim}
	\mathcal{L}_{NCS}(\mathbf{z}^{1}, \mathbf{z}^{2}) = -\frac{\texttt{Pred}(\mathbf{z}^{1})}{\lVert\texttt{Pred}(\mathbf{z}^{1})\rVert_{2}}\cdot\frac{\mathbf{z}^{2}}{\lVert\mathbf{z}^{2}\rVert_{2}},
\end{equation}
where $\lVert\cdot\rVert_{2}$ is $\ell_{2}$-norm. Following \cite{Chen21c}, we define a symmetric loss as:
\begin{equation}\label{eqn:loss_rcop_sym}
	\hat{\mathcal{L}}_{RCoP} = \frac{1}{2}\mathcal{L}_{NCS}(\mathbf{z}^{1}, \mathbf{z}^{2}) + \frac{1}{2}\mathcal{L}_{NCS}(\mathbf{z}^{2}, \mathbf{z}^{1}).
\end{equation}
It has been empirically verified that the stop-gradient (\texttt{StopGrad}) operation plays a crucial role in preventing SimSiam from representation collapse \cite{Chen21c}, i.e., using modified loss $\mathcal{L}_{NCS}(\mathbf{z}^{1}, \texttt{StopGrad}(\mathbf{z}^{2}))$ in \eqref{eqn:loss_neg_cos_sim}. This implies that $\mathbf{z}^{2}$ is viewed as a constant therein, and would not receive gradient information from the NCS loss during training. A similar setting holds for $\mathcal{L}_{NCS}(\mathbf{z}^{2}, \texttt{StopGrad}(\mathbf{z}^{1}))$. To this end the loss function of RCoP in \eqref{eqn:loss_rcop_sym} is reformulated as:
\begin{align}
	\mathcal{L}_{RCoP} &= \frac{1}{2}\mathcal{L}_{NCS}(\mathbf{z}^{1}, \texttt{StopGrad}(\mathbf{z}^{2})) \nonumber \\
	                   &{}\quad + \frac{1}{2}\mathcal{L}_{NCS}(\mathbf{z}^{2}, \texttt{StopGrad}(\mathbf{z}^{1})). \label{eqn:loss_rcop_w_stopgrad}
\end{align}

Note that a full discussion of the relationship between stop-gradient and optimization dynamic is beyond the scope of this paper. Readers may refer to \cite{Chen21c} and \cite{Tian21} for more details. Similar to SimSiam, RCoP can learn meaningful audio-visual representations without using negative samples, large batches, and momentum encoders. This simplicity of implementation makes RCoP reduce the demand for large memory space and high performance computing, and hence suitable for processing video data.

\section{Experiments}\label{sec:experiments}
\subsection{Experimental Setup}
\subsubsection{Datasets}
We perform experiments on three music video datasets: MUSIC-11 \cite{Zhao18}, MUSIC-21 \cite{Zhao19}, and URMP \cite{Li18}. MUSIC-11 dataset contains instrument solo and duet videos collected by keyword query from YouTube. There are 11 instrument categories considered therein, namely accordion, acoustic guitar, cello, clarinet, erhu, flute, saxophone, trumpet, tuba, violin, and xylophone. The original dataset has 565 videos of solos and 149 videos of duets, but about 11$\%$ of which have been removed by the YouTube users at the time of conducting experiments. For a fair comparison, we replaced the unavailable entries with similar YouTube videos, and finally yielding 516 solo and 143 duet videos, respectively. Following \cite{Gao19} we select the first and second videos in each instrument category to construct validation and test datasets, and the rest ones as training data. All videos are split into 20s clips. MUSIC-21 dataset is an extended version of MUSIC-11 and is more challenging for the task of visual sound separation. In addition to the above 11 instrument categories it also includes another 10 categories, i.e., bagpipe, banjo, bassoon, congas, drum, electric bass, guzheng, piano, pipa, and ukulele. The original MUSIC-21 dataset has 1365 untrimmed videos of musical solos and duets, however the number of available terms used in our experiments is 1226. The data split method for this dataset is the same as MUSIC-11. URMP dataset comprises totally 44 multi-instrument musical pieces recorded in studio. We use duet videos from this dataset as part of test samples in the qualitative comparison experiments.

\subsubsection{Implementation Details}
For the visual data pre-processing, we extract video frames at 8 FPS and perform data augmentation on each frame with random scaling, cropping, and horizontal flipping at training stage, like \cite{Zhao18,Xu19,Zhu20a}. Unless otherwise indicated, we only use 3 frames per 6-second video clip to compute the visually-guided feature map, the same as in \cite{Zhao18}. We also leverage the method described in \cite{Zhao18} to pre-process audio data. Specifically, we first sub-sample each sound signal at 11kHz and then randomly crop a 6-second audio clip for training and test. By using STFT with a Hanning window size of 1022 and a hop length of 256, we transform the input audio into a time-frequency (T-F) spectrogram of size $512\times256$. The spectrogram is further re-sampled on a log-frequency scale to produce a T-F representation of size $256\times256$.

\begin{table*}[t]
	\renewcommand\arraystretch{1.2}
	\caption{Sound source separation results on a held out MUSIC-11 test set ($N = 2$ mixture). The ``repro.'' corresponds to results of our reproduction, while results of other compared methods are from \cite{Gao19,Hu20,Qian20}. Numbers annotated with $^{*}$ denote estimations based on network structures described in corresponding papers. Best results are in \textbf{bold}, and second best ones are \uline{underlined}.}
	\label{tab:separate_results_music11}
	\centering
	\begin{tabular}{l|S[table-format=3.2]|S[detect-weight, table-format=1.2]S[detect-weight, table-format=2.2]S[detect-weight, table-format=2.2]|S[detect-weight, table-format=1.2]S[detect-weight, table-format=2.2]S[detect-weight, table-format=2.2]}
		\hline
		\multirow{2}*{Method}          			&{\multirow{2}*{\#Params (M)}}  &\multicolumn{3}{c|}{2-Mix}                          &\multicolumn{3}{c}{3-Mix}                           \\ \cline{3-8}
		&{}         				   	&{SDR $\uparrow$}&{SIR $\uparrow$}&{SAR $\uparrow$}  &{SDR $\uparrow$}&{SIR $\uparrow$}&{SAR $\uparrow$}  \\
		\hline\hline
		NMF-MFCC \cite{Spiertz09}      			&\ \ N/A    			 	    &0.92            &5.68	   	      &6.84  	  		 &0.92		      &5.68	           &6.84	  		  \\
		AV-Mix-Sep \cite{Gao19}        			&71.65       				    &3.16         	 &6.74	          &8.89  	  		 &3.23		      &7.01	           &9.14	  		  \\
		Qian's \cite{Qian20}		   			&58,96   					 	&6.53          	 &12.15	    	  &11.31	  		 &6.57   	      &11.90	       &10.78    		  \\
		CAVL \cite{Hu20}			   			&85.36$^{*}$  					&6.59         	 &10.10	    	  &12.56	  		 &6.78   	      &10.62	       &12.19    		  \\
		Sound-of-Pixels \cite{Zhao18}   		&41.51   					 	&7.30         	 &11.90        	  &11.90      		 &6.05            &9.81            &12.40    		  \\
		Sound-of-Pixels (repro.) \cite{Zhao18}  &41.51   					 	&7.90         	 &12.08	          &11.92      		 &5.97   	      &9.65	           &11.22     		  \\
		MP-Net (repro.) \cite{Xu19}             &71.75  					 	&8.05         	 &11.77	          &11.93	      	 &6.24   	      &10.00	       &10.97     		  \\
		Co-Separation \cite{Gao19}	   			&107.88				 	        &7.38         	 &\bfseries 13.70 &10.80	  		 &7.64		      &\uline{13.80}   &11.30	  		  \\
		AVPC (Ours)				       			&16.16   					 	&\uline{8.54}  	 &12.85	          &\uline{12.93}	 &\uline{8.56}    &13.06	       &\uline{12.53}     \\
		AVPC-RCoP (Ours)		       			&16.16   					 	&\bfseries 9.26  &\uline{13.60}   &\bfseries 12.97	 &\bfseries 9.33  &\bfseries 14.09 &\bfseries 13.04   \\
		\hline
	\end{tabular}
\end{table*}

We adopt the AdamW \cite{Loshchilov19} optimizer to train our AVPC model, where the weight decay coefficient is 1e-2, and the step size hyperparameter $\beta_{1}$ is equal to 0.9 and $\beta_{2}$ is 0.999 in all cases. The sound separation network uses a learning rate of 1e-3; all parameters of video analysis network are froze, except for the additional convolutional layer concatenated with the backbone structure, which uses a learning rate of 1e-4. In case of learning audio-visual representation with RCoP (i.e., AVPC-RCoP), we employ the SGD optimizer at first training stage (i.e., training with RCoP), where the momentum is 0.9, weight decay is 1e-4, and the predictor uses a learning rate of 1e-3; the remaining settings at second training stage (i.e., training with MaS) are the same as the case of AVPC. To stabilize the learning process, we utilize the batch normalization variant \cite{Laurent16} at each layer and at each time step in PCNet, except the visual feature prediction at bottom layer.

\subsubsection{Evaluation Metrics}
Because in MUSIC dataset the ground truth sound components of real videos containing multiple sounds are unknown, we use the synthetic mixture audios for quantitative evaluation, similar to \cite{Zhao18,Gao18,Owens18,Zhao19,Gao19,Gan20,Zhu20a}. Three metrics are adopted to quantify performance: Signal-to-Distortion Ratio (SDR), Signal-to-Interference Ratio (SIR), and Signal-to-Artifact Ratio (SAR). SIR indicates the suppression of interference; SAR reflects the artifacts introduced by the separation process; and SDR measures the overall performance \cite{Huang14}. Their units are in dB, and higher is better for all three metrics.

\begin{table}[t]
	\scriptsize
	\renewcommand\arraystretch{1.2}
	\caption{Sound source separation results on a held out MUSIC-21 test set ($N = 2$ mixture). The separation scores of all compared methods are from \cite{Zhao19}, except those of ``repro.'' methods.}
	\label{tab:separate_results_music21_2mix}
	\centering
	\begin{tabular}{l|S[table-format=2.2]|S[detect-weight, table-format=1.2]S[detect-weight, table-format=2.2]S[detect-weight, table-format=2.2]}
		\hline
		Method          						&{\#Params (M)}  	&{SDR $\uparrow$}&{SIR $\uparrow$}&{SAR $\uparrow$}  \\		   						
		\hline\hline
		NMF \cite{Virtanen07}          			&\ N/A   			&2.78            &6.70	   	      &9.21  	  		 \\
		Deep-Separation \cite{Chandna17}    	&9.99$^{*}$      	&4.75         	 &7.00	          &10.82  	  		 \\
		MIML \cite{Gao18}			   			&60.19    			&4.25         	 &6.23	    	  &11.10	  		 \\
		Sound-of-Pixels \cite{Zhao18}  			&41.60   			&7.52         	 &13.01        	  &11.53      		 \\
		Sound-of-Pixels (repro.) \cite{Zhao18}  &41.60   			&8.47         	 &12.45	          &12.27	      	 \\
		MP-Net (repro.) \cite{Xu19}             &71.84   			&7.14         	 &10.31	          &12.13      		 \\
		Sound-of-Motions \cite{Zhao19}	    	&63.23$^{*}$  		&8.31         	 &\bfseries 14.82 &\uline{13.11}	 \\
		AVPC (Ours)				      	 		&16.16   			&\uline{8.91}    &12.99	          &13.07	      	 \\
		AVPC-RCoP (Ours)		       			&16.16   			&\bfseries 9.50  &\uline{13.74}   &\bfseries 13.29   \\
		\hline
	\end{tabular}
\end{table}
\subsection{Quantitative Results}\label{sec:quantitative_exp}
We first evaluate the method performance in the task of separating sounds from two different kinds of instruments. On MUSIC-11 we train our model with two different data sources, i.e., 2-Mix and 3-Mix, which contain 2 and 3 sound components in the mixture, respectively. We also use the publicly released code to retrain two related models, Sound-of-Pixels\footnote{\href{https://github.com/hangzhaomit/Sound-of-Pixels}{https://github.com/hangzhaomit/Sound-of-Pixels}} \cite{Zhao18} and MP-Net\footnote{\href{https://github.com/SheldonTsui/Minus-Plus-Network}{https://github.com/SheldonTsui/Minus-Plus-Network}} \cite{Xu19}. As shown in Table \ref{tab:separate_results_music11}, performance of pure sound separation approach NMF-MFCC \cite{Spiertz09} is poorer than visually-guided methods, indicating the benefit of leveraging visual features to separate sounds. By employing an object detection network pre-trained on other large-scale dataset, Co-Separation \cite{Gao19} reports superior performance over other methods that use only instrument appearance information. However, its superiority comes at the cost of increased model size (cf., the biggest \#Param, 107.88M, among all compared methods). By comparison, AVPC enables separation quality on par with or better than the state-of-the-art on MUSIC-11, while reducing a significant number of network parameters (cf., the smallest \#Param, 16.16M, among all compared methods). The results indicate that the PC-based sound separation network is essential. Besides, AVPC-RCoP further improves AVPC's performance especially on SDR and SIR metrics, demonstrating that the learning strategy RCoP can provide reasonable parameter initialization for the audio-visual model. What's more, both Co-Separation and ours consistently outperform all other competitors across two training configurations (i.e., 2-Mix and 3-Mix), and thus show better generalization ability on this dataset.

\begin{table}[t]
	\renewcommand\arraystretch{1.2}
	\caption{Sound separation performances with $N=3,4$ mixtures on a held out MUSIC-21 test set.}
	\label{tab:separate_results_music21_34mix}
	\centering
	\begin{tabular}{c|l|S[table-format=1.2]S[table-format=1.2]S[table-format=1.2]}
		\hline
		$N$       			&Method          				            &{SDR $\uparrow$}&{SIR $\uparrow$}&{SAR $\uparrow$}  \\		   						
		\hline\hline
		\multirow{4}*{3}    
		&Sound-of-Pixels (repro.)\cite{Zhao18}  	&4.29            &8.08        	  &8.67      		 \\
		&MP-Net (repro.) \cite{Xu19}	    		&3.14         	 &6.42 	          &8.68	  		     \\
		&AVPC (Ours)				      	 		&4.48         	 &8.86	          &8.69	      		 \\
		&AVPC-RCoP (Ours)		       			    &4.31         	 &8.26	          &8.84	      		 \\
		\hline
		\multirow{4}*{4}    &Sound-of-Pixels (repro.)\cite{Zhao18}		&2.01            &5.21            &7.35      		 \\
		&MP-Net (repro.) \cite{Xu19}	    		&1.54            &4.67 	          &7.61	  		     \\
		&AVPC (Ours)				      	 		&2.61            &6.89	          &7.12	      		 \\
		&AVPC-RCoP (Ours)		       			    &2.20         	 &6.16	          &7.13	      		 \\
		\hline
	\end{tabular}
\end{table}

\begin{figure*}
	\centering
	\resizebox{0.9\textwidth}{!}{%
		\footnotesize
		\begin{tabular}{cc|c|c}
			{}            &Mix. Pair 1 (MUSIC-11)      &Mix. Pair 2 (MUSIC-21)   &Mix. Pair 3 (URMP)   \\
			
			Video Frames &\includegraphics[width=0.1\linewidth]{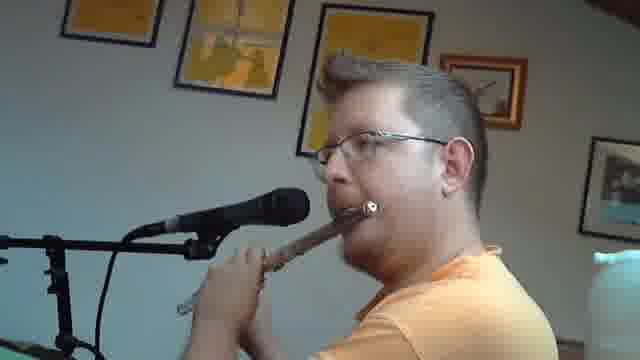}\hspace{0.5em} \includegraphics[width=0.1\linewidth]{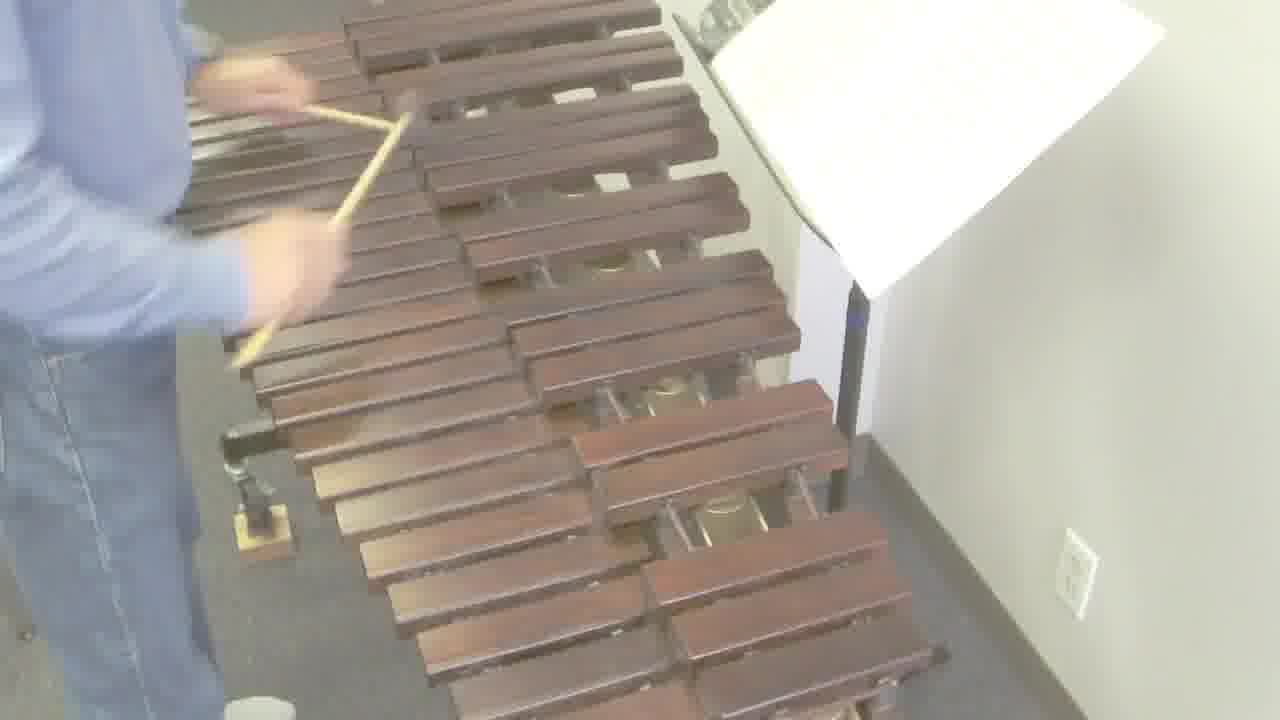}    &\includegraphics[width=0.1\linewidth]{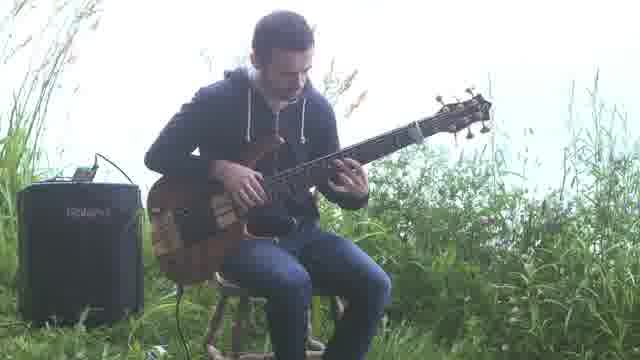}\hspace{0.5em} \includegraphics[width=0.1\linewidth]{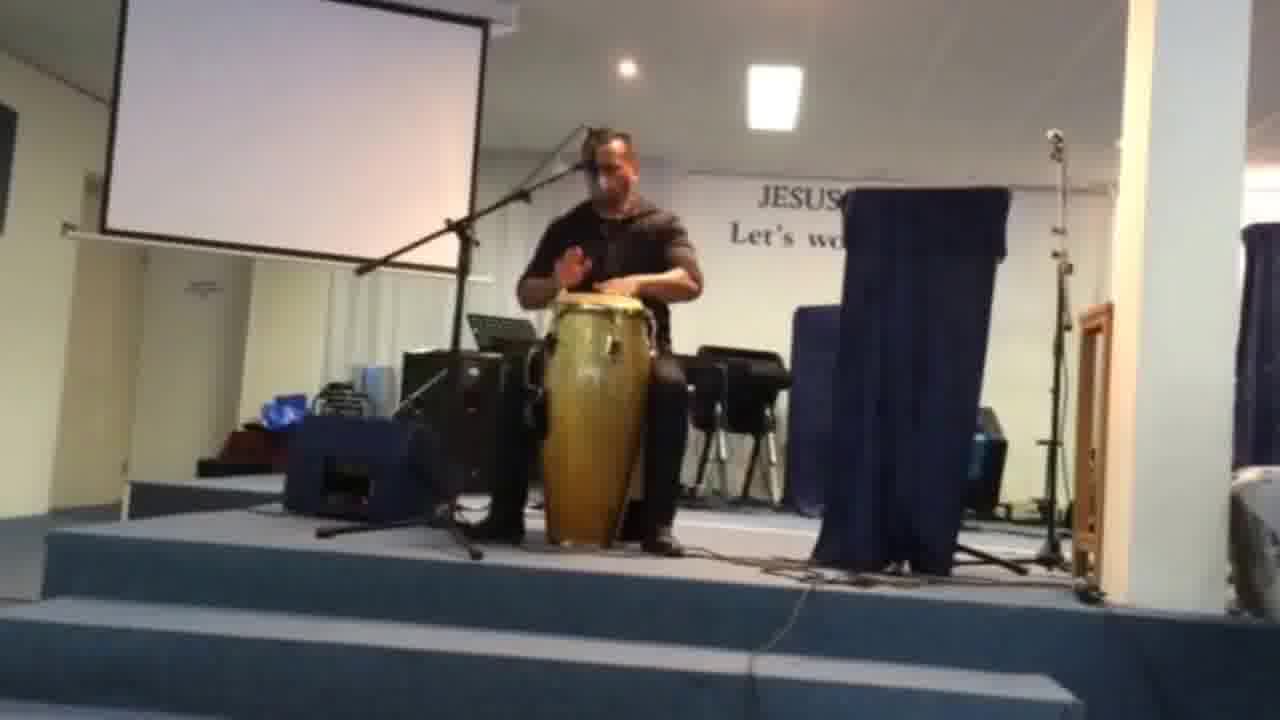}  &\includegraphics[width=0.1\linewidth]{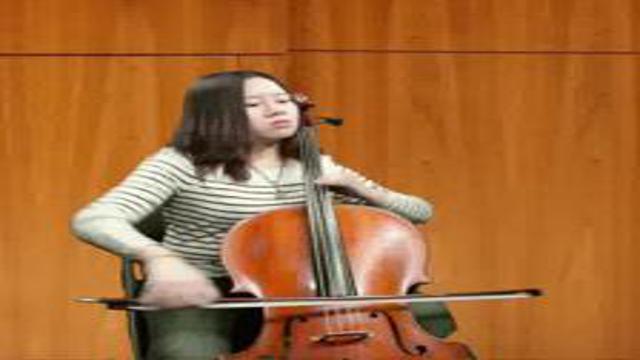}\hspace{0.5em} \includegraphics[width=0.1\linewidth]{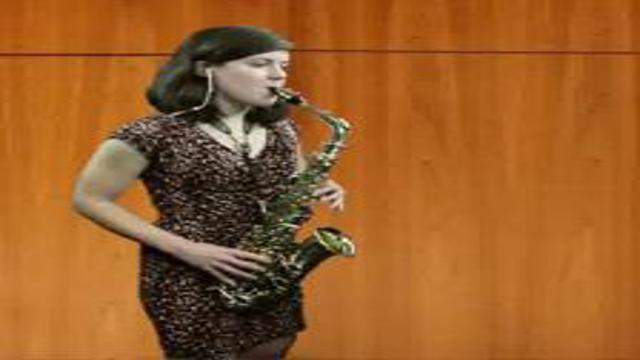}  \\
			
			\multirowcell{Mixture \\ Spectrogram}   &\includegraphics[width=0.1\linewidth]{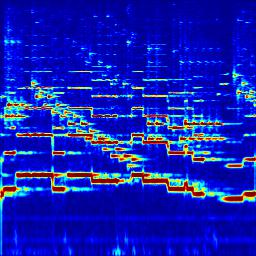}  &\includegraphics[width=0.1\linewidth]{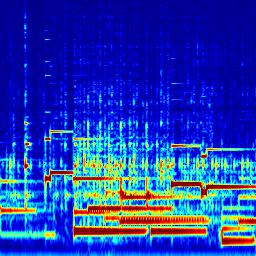}  &\includegraphics[width=0.1\linewidth]{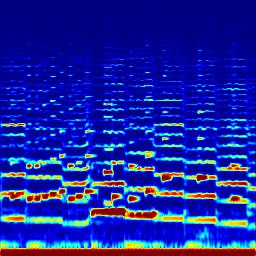}  \\	
			
			\multirowcell{Ground Truth \\ Mask}       &\fbox{\includegraphics[width=0.1\linewidth]{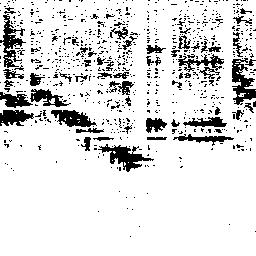}}\hspace{0.5em} \fbox{\includegraphics[width=0.1\linewidth]{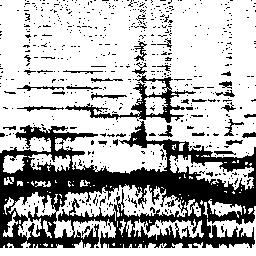}}    &\fbox{\includegraphics[width=0.1\linewidth]{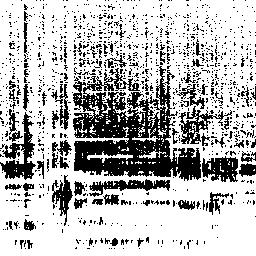}}\hspace{0.5em} \fbox{\includegraphics[width=0.1\linewidth]{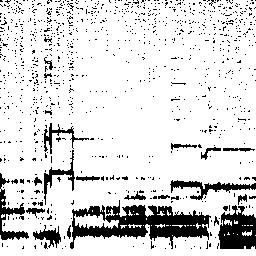}}  &\fbox{\includegraphics[width=0.1\linewidth]{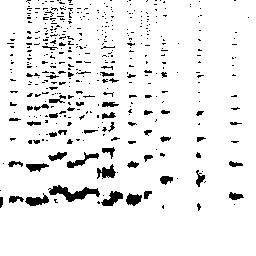}}\hspace{0.5em} \fbox{\includegraphics[width=0.1\linewidth]{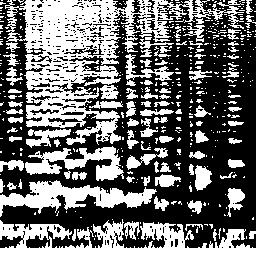}}  \\
			
			\multirowcell{Pred. Mask \\ (Sound-of-Pixels)} &\fbox{\includegraphics[width=0.1\linewidth]{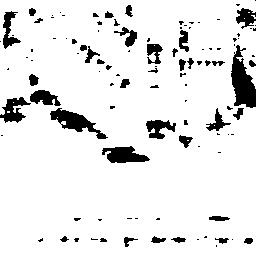}}\hspace{0.5em} \fbox{\includegraphics[width=0.1\linewidth]{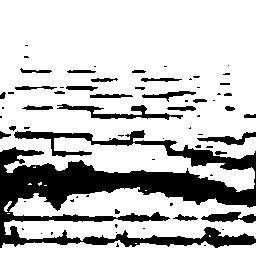}}  
			&\fbox{\includegraphics[width=0.1\linewidth]{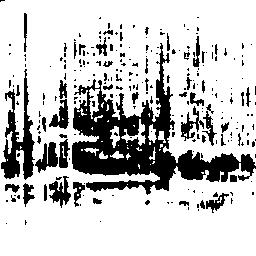}}\hspace{0.5em} \fbox{\includegraphics[width=0.1\linewidth]{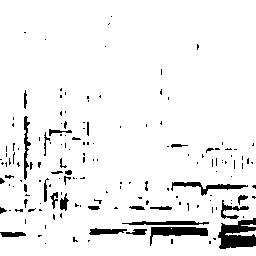}}  &\fbox{\includegraphics[width=0.1\linewidth]{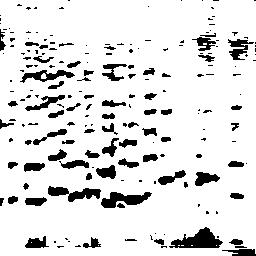}}\hspace{0.5em} \fbox{\includegraphics[width=0.1\linewidth]{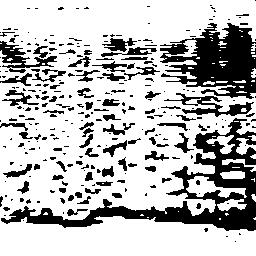}}\\
			
			\multirowcell{Pred. Mask \\ (MP-Net)} &\fbox{\includegraphics[width=0.1\linewidth]{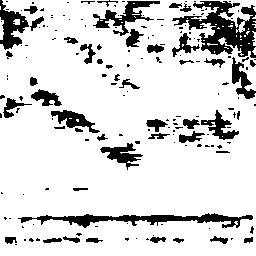}}\hspace{0.5em} \fbox{\includegraphics[width=0.1\linewidth]{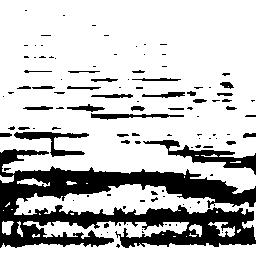}}  &\fbox{\includegraphics[width=0.1\linewidth]{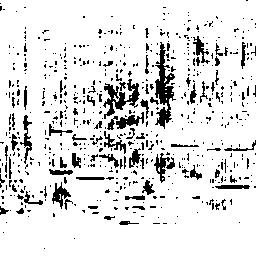}}\hspace{0.5em} \fbox{\includegraphics[width=0.1\linewidth]{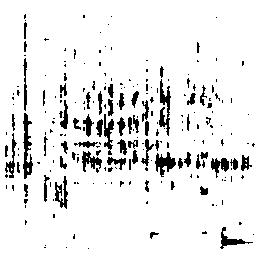}}  &\fbox{\includegraphics[width=0.1\linewidth]{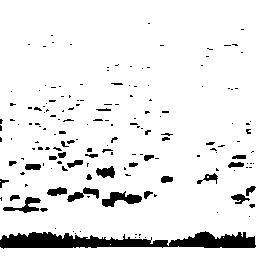}}\hspace{0.5em} \fbox{\includegraphics[width=0.1\linewidth]{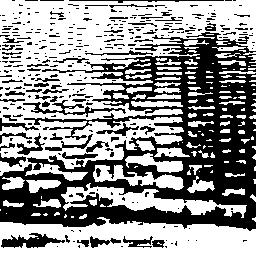}}  \\
			
			\multirowcell{Pred. Mask \\ (AVPC)} &\fbox{\includegraphics[width=0.1\linewidth]{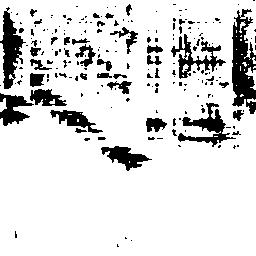}}\hspace{0.5em} \fbox{\includegraphics[width=0.1\linewidth]{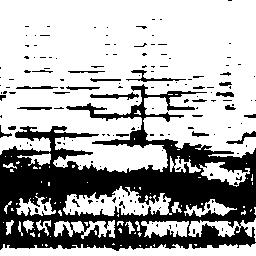}}  &\fbox{\includegraphics[width=0.1\linewidth]{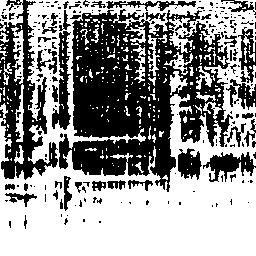}}\hspace{0.5em} \fbox{\includegraphics[width=0.1\linewidth]{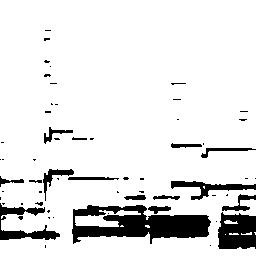}}  &\fbox{\includegraphics[width=0.1\linewidth]{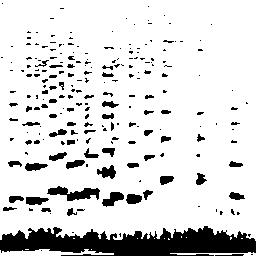}}\hspace{0.5em} \fbox{\includegraphics[width=0.1\linewidth]{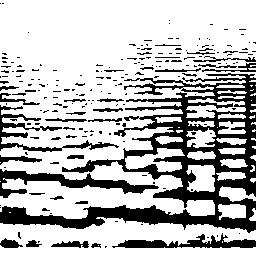}}  \\
			
			\multirowcell{Pred. Mask \\ (AVPC-RCoP)} &\fbox{\includegraphics[width=0.1\linewidth]{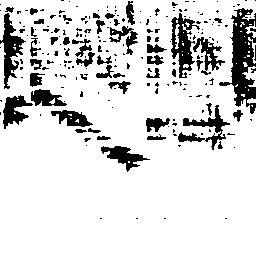}}\hspace{0.5em} \fbox{\includegraphics[width=0.1\linewidth]{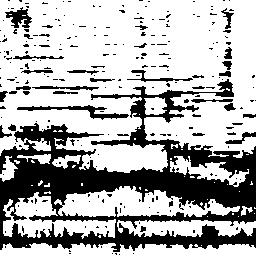}} &\fbox{\includegraphics[width=0.1\linewidth]{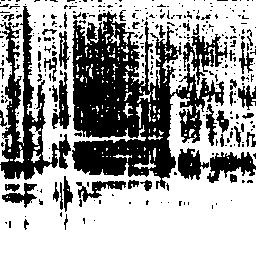}}\hspace{0.5em} \fbox{\includegraphics[width=0.1\linewidth]{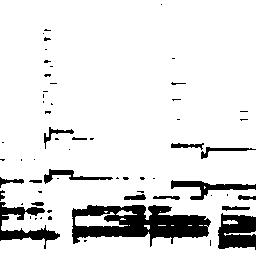}} &\fbox{\includegraphics[width=0.1\linewidth]{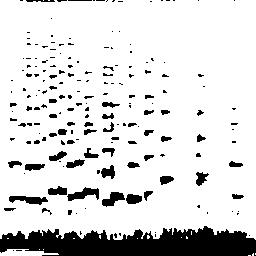}}\hspace{0.5em} \fbox{\includegraphics[width=0.1\linewidth]{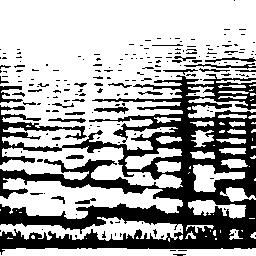}}\\
			
			\multirowcell{Ground Truth \\ Spectrogram} &\includegraphics[width=0.1\linewidth]{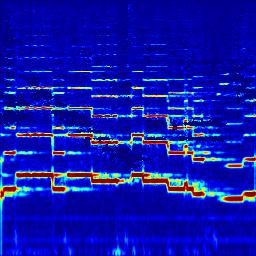}\hspace{0.5em} \includegraphics[width=0.1\linewidth]{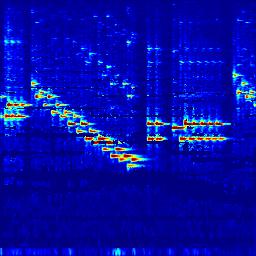}    &\includegraphics[width=0.1\linewidth]{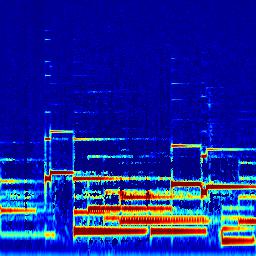}\hspace{0.5em} \includegraphics[width=0.1\linewidth]{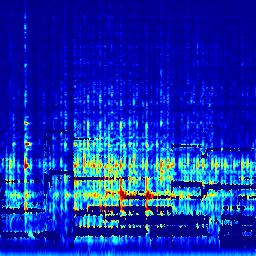}  &\includegraphics[width=0.1\linewidth]{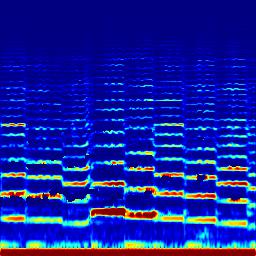}\hspace{0.5em} \includegraphics[width=0.1\linewidth]{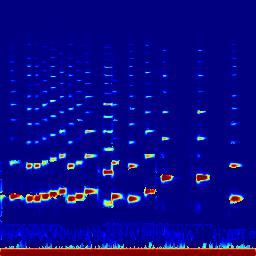}  \\
			
			\multirowcell{Pred. Spectrogram \\ (Sound-of-Pixels)} &\includegraphics[width=0.1\linewidth]{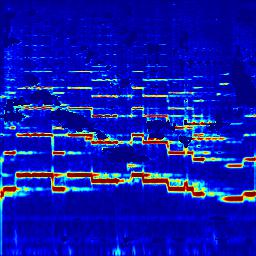}\hspace{0.5em} \includegraphics[width=0.1\linewidth]{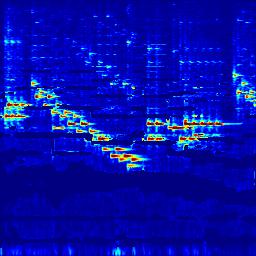}    &\includegraphics[width=0.1\linewidth]{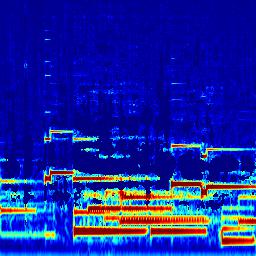}\hspace{0.5em} \includegraphics[width=0.1\linewidth]{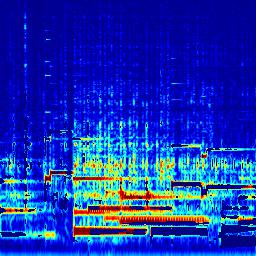}  &\includegraphics[width=0.1\linewidth]{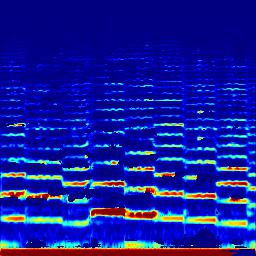}\hspace{0.5em} \includegraphics[width=0.1\linewidth]{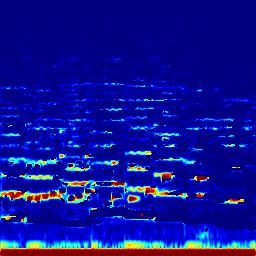}  \\
			
			\multirowcell{Pred. Spectrogram \\ (MP-Net)} &\includegraphics[width=0.1\linewidth]{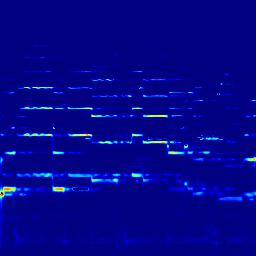}\hspace{0.5em} \includegraphics[width=0.1\linewidth]{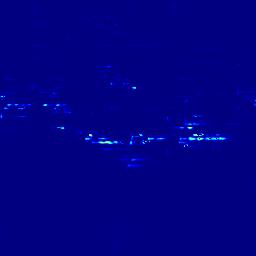}    &\includegraphics[width=0.1\linewidth]{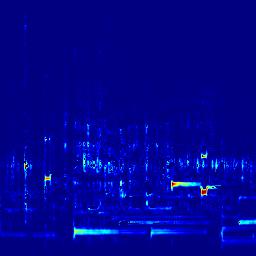}\hspace{0.5em} \includegraphics[width=0.1\linewidth]{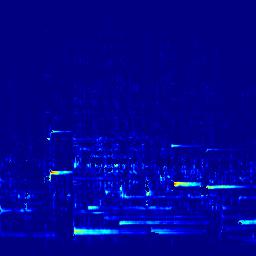}  &\includegraphics[width=0.1\linewidth]{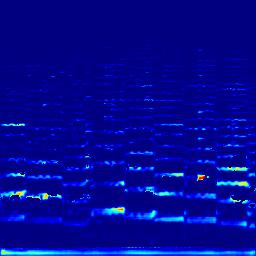}\hspace{0.5em} \includegraphics[width=0.1\linewidth]{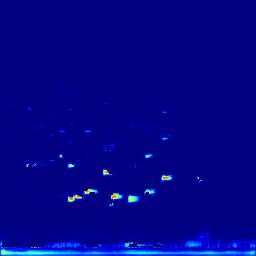}  \\
			
			\multirowcell{Pred. Spectrogram \\ (AVPC)} &\includegraphics[width=0.1\linewidth]{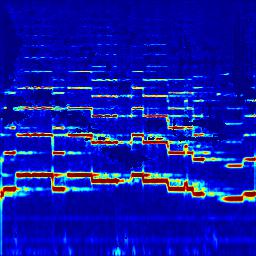}\hspace{0.5em} \includegraphics[width=0.1\linewidth]{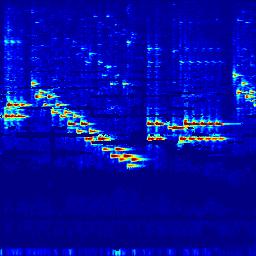}    &\includegraphics[width=0.1\linewidth]{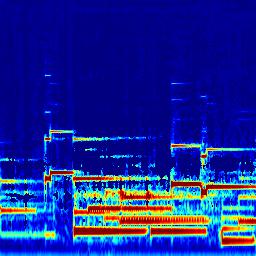}\hspace{0.5em} \includegraphics[width=0.1\linewidth]{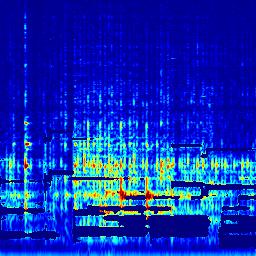}  &\includegraphics[width=0.1\linewidth]{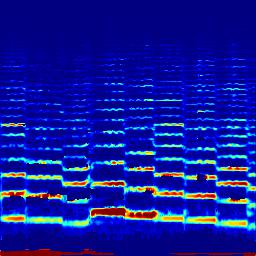}\hspace{0.5em} \includegraphics[width=0.1\linewidth]{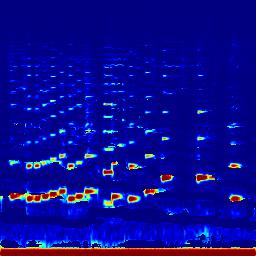}  \\
			
			\multirowcell{Pred. Spectrogram \\ (AVPC-RCoP)} &\includegraphics[width=0.1\linewidth]{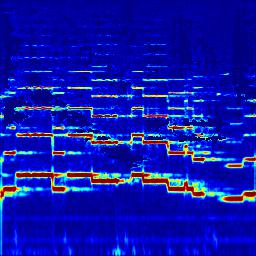}\hspace{0.5em} \includegraphics[width=0.1\linewidth]{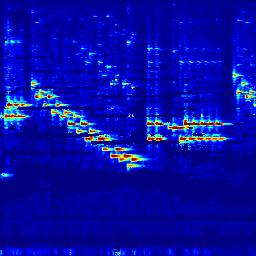}    &\includegraphics[width=0.1\linewidth]{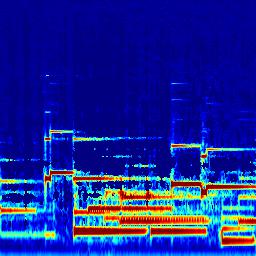}\hspace{0.5em} \includegraphics[width=0.1\linewidth]{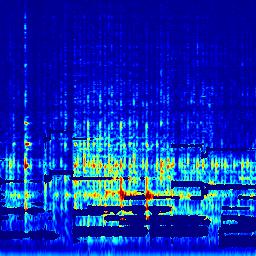}  &\includegraphics[width=0.1\linewidth]{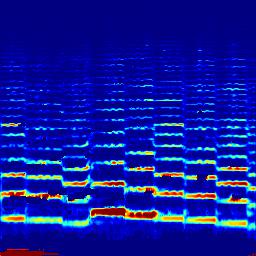}\hspace{0.5em} \includegraphics[width=0.1\linewidth]{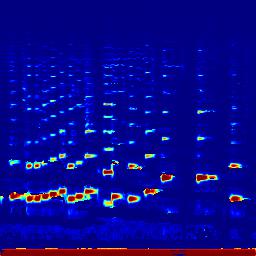}%
	\end{tabular}}
	\caption{Visualization of visual sound separation results, from Sound-of-Pixels \cite{Zhao18}, MP-Net \cite{Xu19}, AVPC, and AVPC-RCoP.}\label{fig:vis_mask_spect}
\end{figure*}

\begin{figure*}
	\centering
	\includegraphics[width=0.95\linewidth]{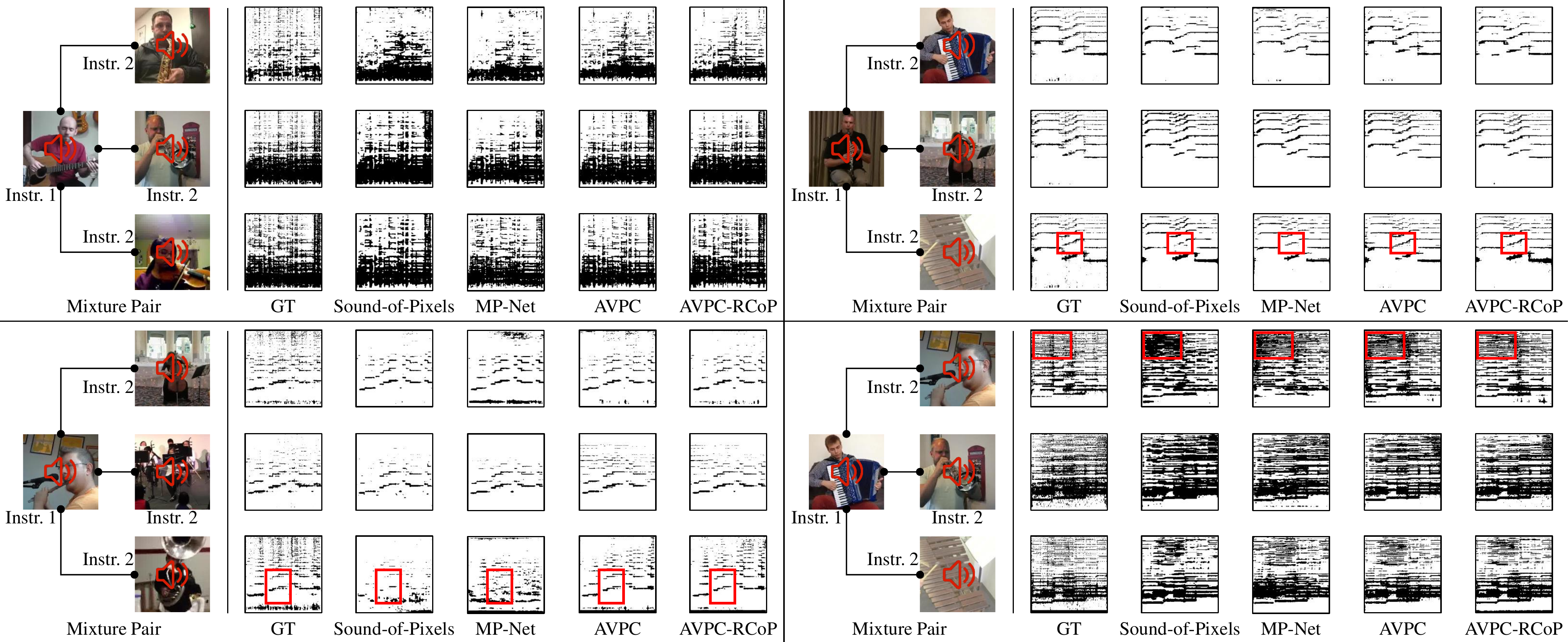}
	\caption{Comparison of predicted masks by separating sounds from mixture audios that share the same one sound component. In each panel, the sound of the first instrument (i.e., Instr. 1) is mixed with various sounds of the second instrument (i.e., Instr. 2), respectively. Here we show sound separation mask used to produce sound of Instr. 2. Note that because the separation mask of Instr. 2 works to filter out sound information about Instr. 1, resultant mask images are visually similar within all mixture pairs that take the same Instr. 1 as one sound component. The visually distinctive mask regions are highlighted with red rectangles. Best viewed by zooming in.}\label{fig:vis_mask_same_comp1}
\end{figure*}

\begin{figure*}[t]
	\centering
	\includegraphics[width=0.48\linewidth]{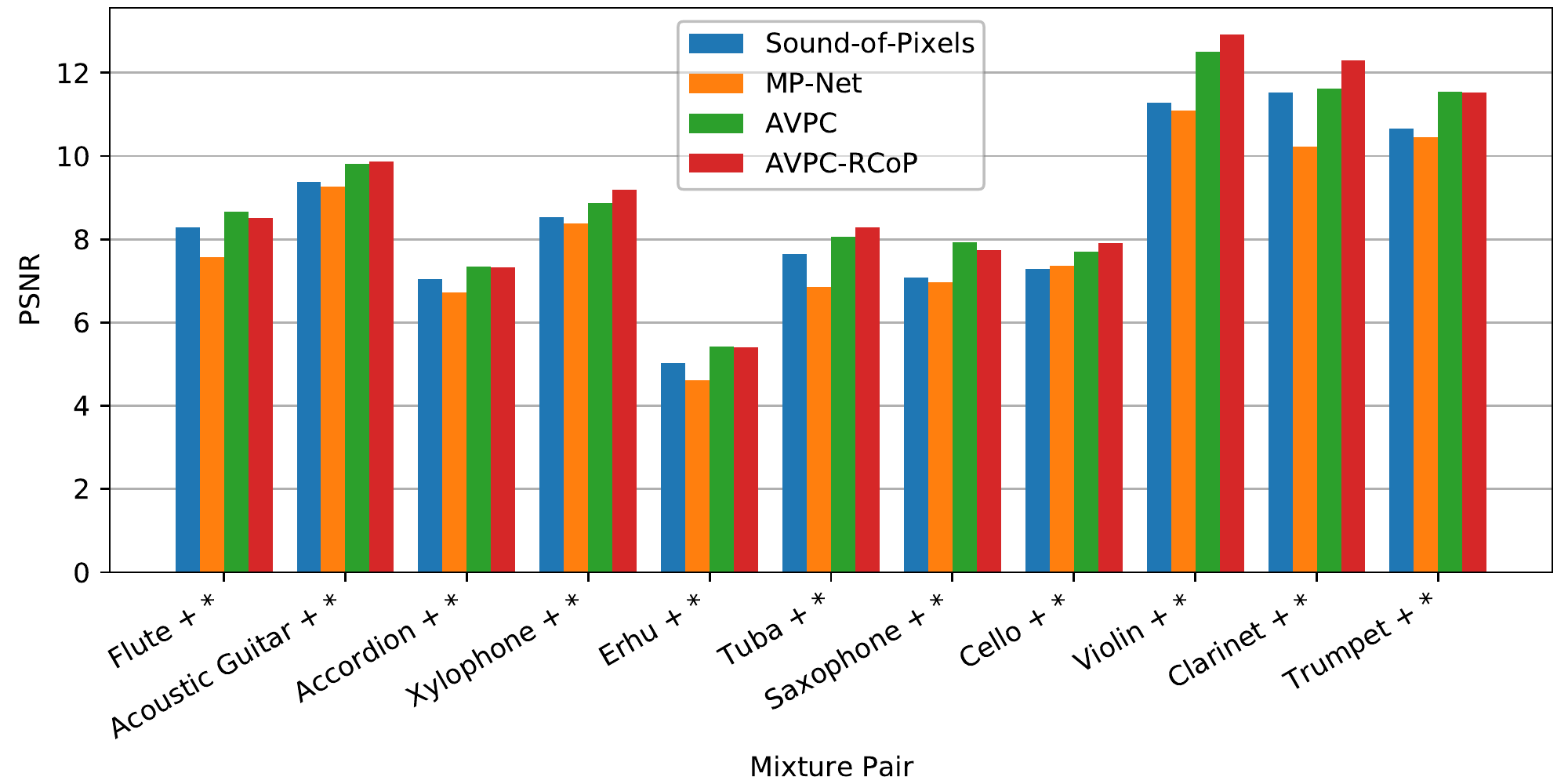}\hspace{0.5em}
	\includegraphics[width=0.48\linewidth]{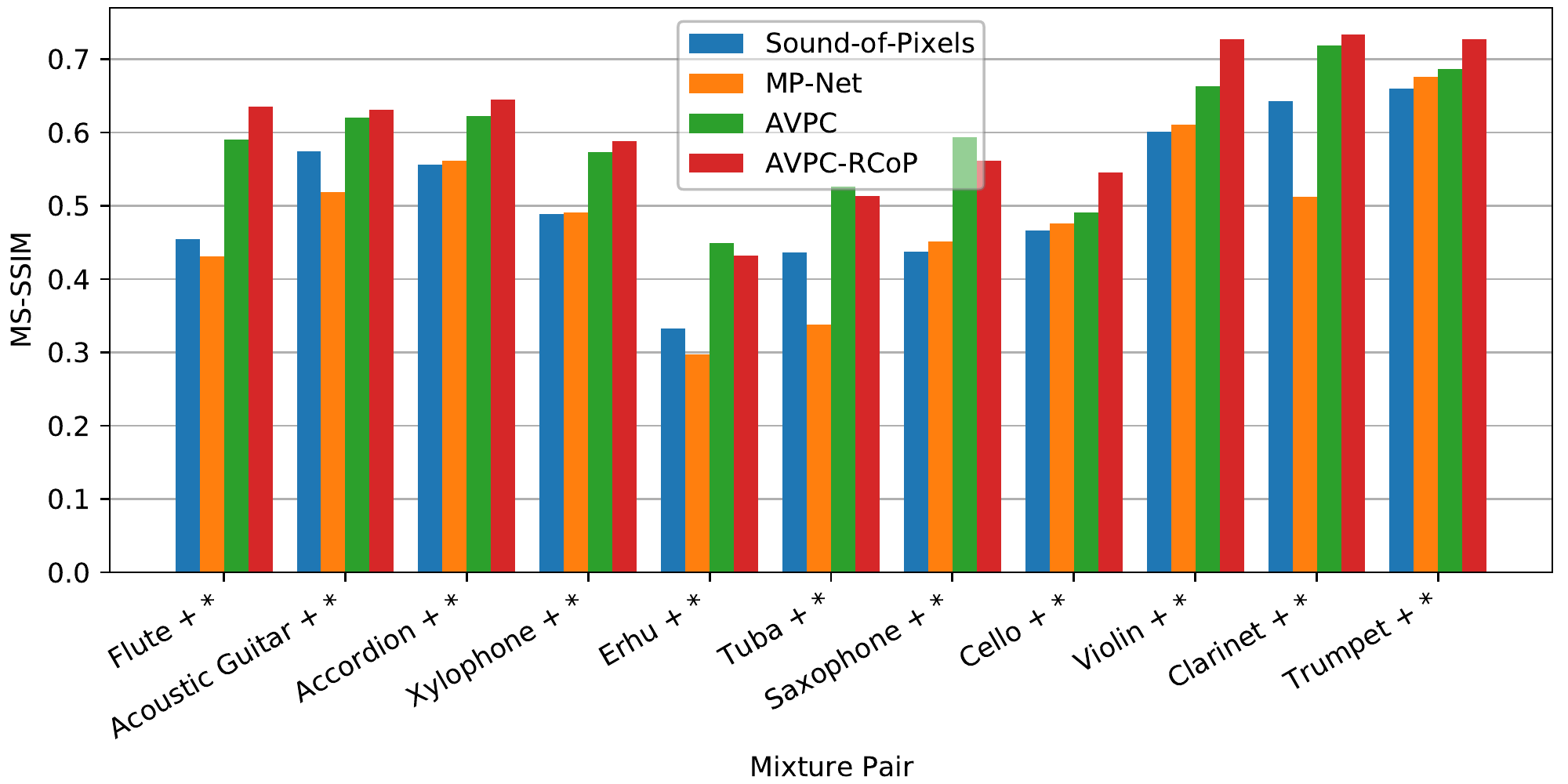}\hspace{0.5em}%
	\caption{Image quality assessment of predicted sound separation masks on MUSIC-11 test set. In each panel, the tick label on horizontal axis ``instrument 1 + *'' denotes all mixture pairs where ``*'' represents instrument 2 that is different from instrument 1. The PSNR (left panel) and MS-SSIM (right panel) quantify average image qualities of the masks used to recover sounds of instrument 2.}\label{fig:iqa}
\end{figure*}

Table \ref{tab:separate_results_music21_2mix} shows the comparison results of separating sounds from two different instruments on MUSIC-21. It can be seen that in such a scenario with more diverse instruments, Sound-of-Motions \cite{Zhao19} performs better than other previous works. Its success roots in taking optical flow and trajectory based motion cues into account, whereas resorting to several existing network models pre-trained on other large-scale datasets. Although the proposed AVPC and AVPC-RCoP, similar to MIML \cite{Gao18}, Sound-of-Pixels, and MP-Net, solely leverage instrument appearance feature to guide sound separation, they obtain top two separation accuracies in terms of SDR, and competitive scores on SIR and SAR, with least parameters among visually-guided methods. This appropriate trade-off between performance and model size can be attributed to the iterative representation inference process in PCNet, which enables AVPC to build increasingly accurate correspondence between audio and visual modalities. We conduct detailed ablation studies to illustrate this rationality in Section \ref{sec:ablation_study}. Additionally, the learning strategy RCoP again improves separation qualities of AVPC across all metrics, showing effectiveness in learning task-oriented audio-visual representations.

Moreover, to explore the behavior of AVPC to address more complex sound separation tasks, we perform additional experiments where the mixture audio consists of 3 or 4 sound components. Here we use parameters of AVPC that were well trained on MUSIC-21 as model initialization, and retrain the network in a simple curriculum learning manner (i.e., training model first on the 3-Mix data source and then on the 4-Mix data source). The two baselines, Sound-of-Pixels and MP-Net, are retrained with the same data and the same training pipeline accordingly. Results are reported in Table \ref{tab:separate_results_music21_34mix}. We find that in these highly mixed cases our AVPCs are still superior to baseline methods. In particular, at $N=4$ the SDR of AVPC outperforms that of Sound-of-Pixels by 0.6dB, and that of MP-Net by 1.07dB. The results manifest that AVPC can generalize better over two baselines. However, the performance of all approaches degrades as increasing the number of sound components in test mixture audios, indicating that separating sound mixtures with multiple sources remains a very challenging task. We leave this to future work.

\begin{table*}[t]
	\renewcommand\arraystretch{1.3}
	\caption{Ablation study of AVPC. ``Iterative Inference'' indicates the iterative representation inference procedure in \eqref{eqn:update_p_fb_avpc}-\eqref{eqn:update_r_ff_avpc}, where we set $T=5$. ``Transform Visual Feature'' denotes performing additional convolution on features extracted by visual backbone networks. ``More Frames'' represents feeding all 48 frames of a 6-second video clip into video analysis network.}
	\label{tab:ablation_study}
	\centering
	\begin{tabular}{c|ccc|cc|S[table-format=1.2]S[table-format=2.2]S[table-format=2.2]}
		\hline
		\multirow{2}*{Model} &\multicolumn{3}{c|}{Training Stage}                     &\multicolumn{2}{c|}{Test Stage}   &{\multirow{2}*{SDR $\uparrow$}}&{\multirow{2}*{SIR $\uparrow$}}&{\multirow{2}*{SAR $\uparrow$}}\\
		\cline{2-6}
		&Iterative Inference&Transform Visual Feature&RCoP       &Iterative Inference &More Frames  &{}                             &{}                             &{}                         \\
		\hline\hline
		A                    &{}                 &{}					  &{}		  &{}                  &{}           &7.01                           &11.90                          &11.09                       \\
		B                    &\checkmark         &{}					  &{}		  &{}                  &{}           &-5.30                          &3.93                           &-2.39                      \\
		C                    &\checkmark         &{}					  &{}		  &\checkmark          &{}           &8.45                           &12.80                          &12.33                       \\
		D                    &\checkmark         &\checkmark              &{}		  &\checkmark          &{}           &8.54                           &12.85                          &12.93                      \\
		E                    &\checkmark         &\checkmark              &\checkmark &\checkmark          &{}           &9.26                           &13.60                          &12.97                      \\
		F                    &\checkmark         &\checkmark              &\checkmark &\checkmark          &\checkmark   &9.40                           &13.78                          &13.06                       \\
		\hline
	\end{tabular}
\end{table*}

\subsection{Qualitative Results}
Fig. \ref{fig:vis_mask_spect} shows qualitative samples with sound separation masks and spectrograms predicted by Sound-of-Pixels, MP-Net, AVPC, and AVPC-RCoP, respectively. As depicted in the mixture pair 1 from MUSIC-11 (i.e., the two columns on the left in Fig. \ref{fig:vis_mask_spect}), Sound-of-Pixels generates masks that allow more pixel information of the mixture sepctrogram to leak out, leading to predicted spectrograms carrying a few tangled frequency features. By contrast, MP-Net excels at filtering out noisy components from predicted spectrograms, but simultaneously removing meaningful details. Compared with these methods, AVPC and AVPC-RCoP produce more accurate sound separation masks and thus spectrograms. Similar phenomena can be observed in mixture pair samples from MUSIC-21 and URMP dataset. Again, these results indicate the superiority of AVPCs in separating musical instrument sounds. More comprehensive results can be found in the supplementary video, where we also show exemplars of separating realistic duet sounds by our AVPCs.

To illustrate the performance stability of AVPC, we conduct experiments to separate sounds from mixture audios that share the same one sound component. Specifically, for each instrument category in MUSIC-11 we select a 6-second audio clip as the first sound component (denoted as Instr. 1), and then synthesize 10 mixture audios by summing the first sound component with the second sound component (denoted as Instr. 2) from the other 10 instrument classes, respectively. The goal is to predict masks that will be used to separate sounds of Instr. 2. Fig. \ref{fig:vis_mask_same_comp1} visualizes four groups of mask prediction examples. We find that Sound-of-Pixels and MP-Net are prone to produce masks underestimating or overestimating occlusion pixels in ground truths (e.g., see marked red regions in top right panel and in bottom right panel in Fig. \ref{fig:vis_mask_same_comp1}, respectively). Compared with two baseline methods, AVPC and AVPC-RCoP exhibit better response to recover local textures in predicted masks. In fact, we also employ two image quality assessment indexes, peak signal-to-noise ratio (PSNR) and multi-scale structure similarity (MS-SSIM) \cite{Wang03}, to evaluate the predicted mask images' quality. From Fig. \ref{fig:iqa}, we observe that both AVPC and AVPC-RCoP produce higher PSNR and MS-SSIM values across all 11 instrument categories than baseline approaches. These results show that for various kinds of instrument mixture audios, AVPC performs consistently well in predicting sound separation masks, hence resulting in sound separation ability superior to Sound-of-Pixels and MP-Net.

\subsection{Ablation Study}\label{sec:ablation_study}
\subsubsection{Effect of Each Part}
To demonstrate how different parts of AVPC influence sound separation performance, we present extensive ablations and analysis in this section. All the results, as summarized in Table \ref{tab:ablation_study}, are based on the held out MUSIC-11 test set. Here, the baseline (model A) is set to learn in a concise way and gets separation scores for reference. By increasing the recursive cycles at training stage ($T=5$) rather than at test stage ($T=1$), we obtain lowest values of the three performance metrics (model B). However, as we also set $T=5$ at test stage (model C), separation scors are significantly improved (compared with model A, model C improves SDR by 1.44dB, SIR by 0.9dB, and SAR by 1.24dB, respectively). This verifies the effectiveness of the iterative inference procedure adopted in AVPC, which only works when the recursive cycles are consistent across traing and test phases. Then, transforming the visual feature through an additional convolutional layer improves performance further (model D). We speculate that the original features derived from off-the-shelf visual backbones (here is ResNet-18) include task-irrelevant noise, and the additional convolutional layer can serve to suppress those noisy information. Moreover, a great improvement is gained after using RCoP during training (model E), showing that RCoP is able to provide a better parameter initialization for the MaS training framework. Ideally AVPC should benefit from sufficient video frames. This is validated by model F that utilizes all 48 frames (corresponding to a 6-second video clip) to extract visual features at test stage.

\subsubsection{Effect of Iterative Representation Inference}\label{sec:ablation_iterative_infer}
\begin{table}[t]
	\renewcommand\arraystretch{1.2}
	\caption{Sound separation performance of AVPC with different recursive cycles at training stage. Results are reported on the held out MUSIC-11 test set ($N = 2$ mixture).}
	\label{tab:perform_vs_cycle_train}
	\centering
	\begin{tabular}{l|S[detect-weight, table-format=2.2]S[detect-weight, table-format=2.2]S[detect-weight, table-format=2.2]S[detect-weight, table-format=2.2]S[detect-weight, table-format=2.2]}
		\hline
		$T$          		&{1}  	  &{3}      	 	&{5}      		  &{6}      		 &{7}           \\		   						
		\hline\hline
		{SDR $\uparrow$}    &6.32     &7.88             &\bfseries 8.54   &8.19   		     &\uline{8.33}  \\
		{SIR $\uparrow$}    &10.49    &\uline{12.70}    &\bfseries 12.85  &12.67  		     &12.67	        \\
		{SAR $\uparrow$}    &11.22    &12.24    		&\uline{12.93}    &\bfseries 12.94   &12.77	        \\		
		\hline
	\end{tabular}
\end{table}

\begin{figure}
	\centering
	\includegraphics[width=0.47\linewidth]{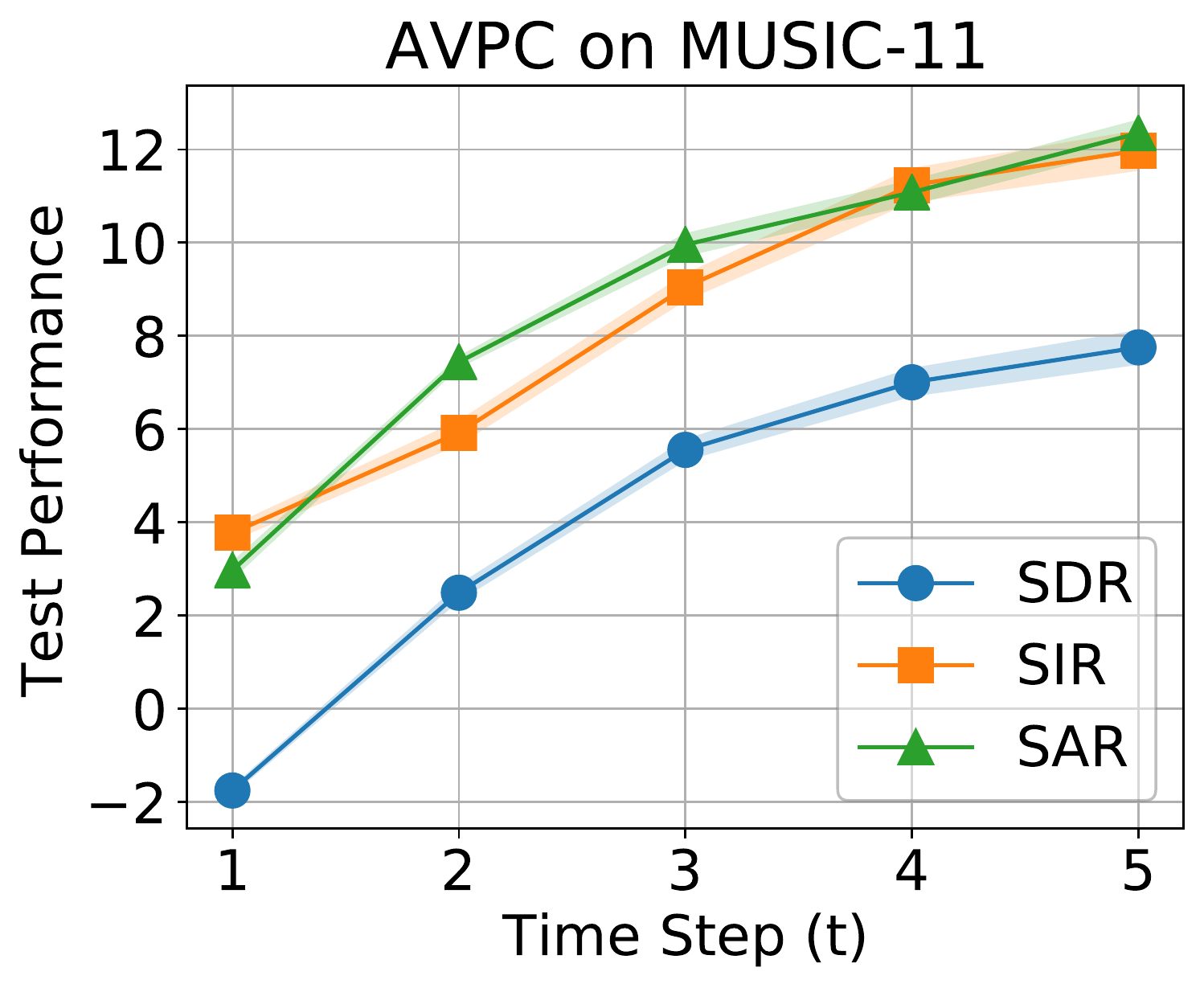}\hspace{1.2em}
	\includegraphics[width=0.47\linewidth]{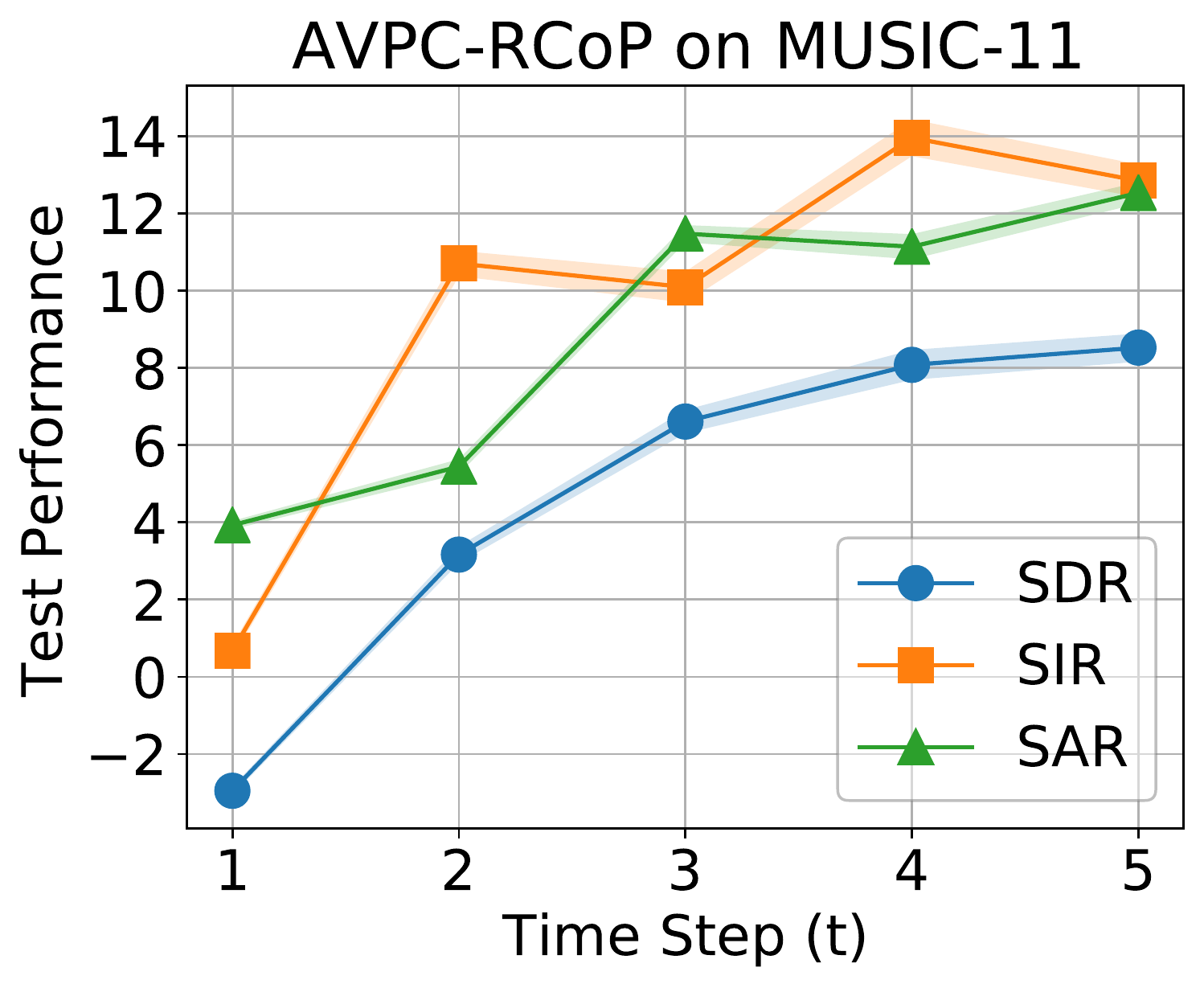}\\
	\caption{Sound separation performance of AVPCs with varying numbers of iterative computing at test stage. The separation scores are derived by running test experiments with 10 random seeds, respectively, and we report the average performance at each time step.}
	\label{fig:perform_vs_cycle_test}
\end{figure}

\begin{figure}
	\centering
	\includegraphics[width=0.95\linewidth]{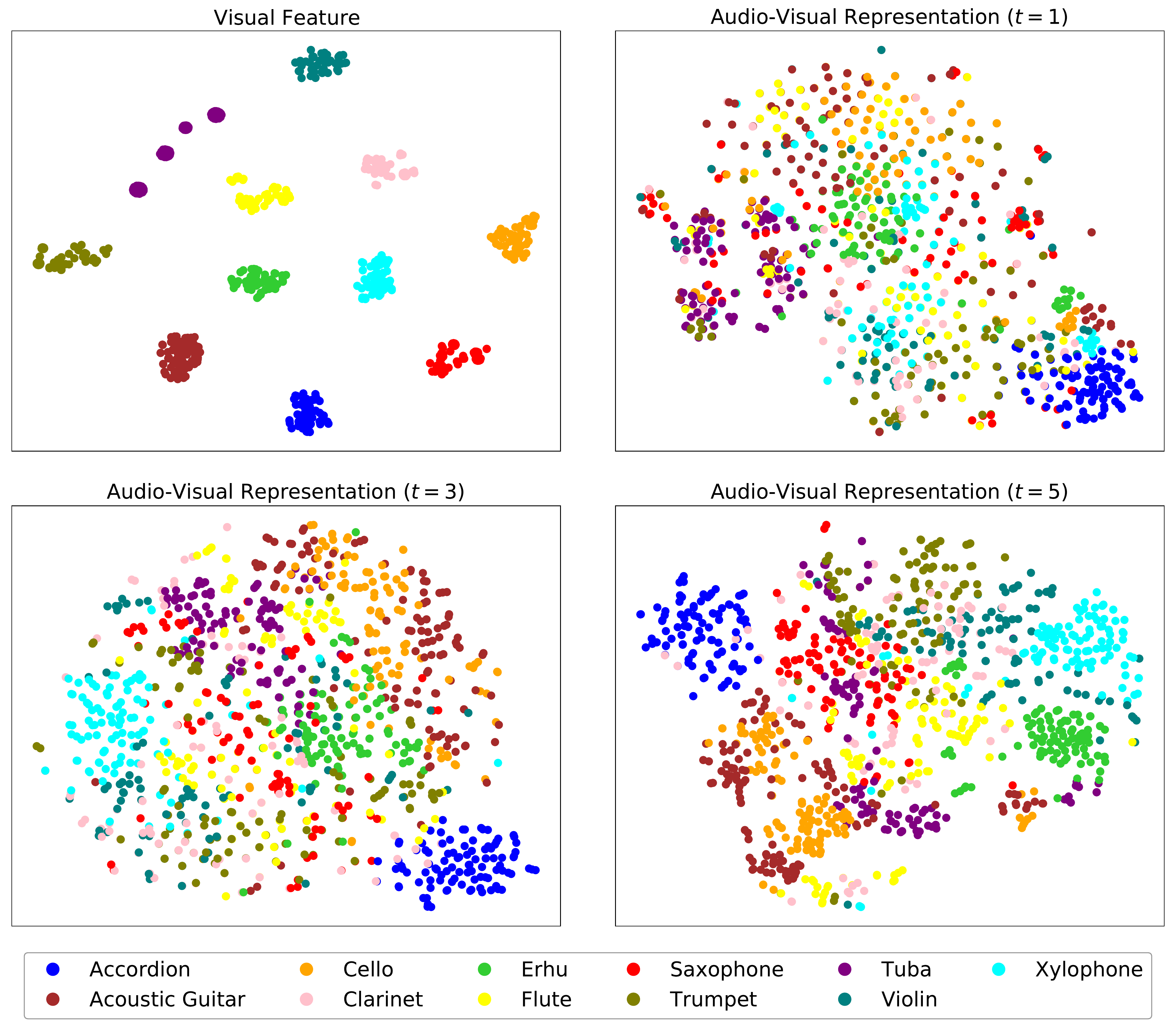}
	\caption{Learned embeddings visualized by t-SNE on MSUIC-11 test set.}
	\label{fig:tsne_fusion_w_time_step}
\end{figure}

The most remarkable nature of AVPC, that distinguishes it from all other audio-visual methods, is inferring representation in an iterative manner. To have an in-depth analysis of this mechanism, we first investigate the effect of different recursive cycles ($T$) at \emph{training} stage. As shown in Table \ref{tab:perform_vs_cycle_train}, AVPC can benefit from training with bigger recursive cycles, however, the performance tends to be saturated when $T>5$. Considering that long-time iterative inference also brings increased computational complexity, we set $T=5$ during training in all experiments. Second, we inspect sound separation results of AVPCs at different computation time steps ($t$) during \emph{test}. As illustrated in Fig. \ref{fig:perform_vs_cycle_test}, all separation scores tend to increase given more cycles of computation, meaning that audio-visual representation is progressively refined by the iterative procedure. In terms of the overall performance metric SDR, AVPC acts slightly better than AVPC-RCoP at initial time step ($t=1$); however, AVPC-RCoP turns out to be superior to AVPC as time goes on, especially at the end of the iterative procedure ($t=5$). This phenomenon reveals that the learning strategy RCoP gains more from increasing the number of iterative computing. Third, to visualize that the iterative inference in AVPC indeed makes audio-visual representation achieve better feature fusion, Fig. \ref{fig:tsne_fusion_w_time_step} displays the t-SNE \cite{Van08} embeddings of visual feature and audio-visual representation, respectively. With the guidance from discriminative visual cues, the audio-visual representations of different sounding objects become more and more distinguishable as increasing time steps (e.g., the embedding cluster of ``Erhu'' denoted by {\color{green}green} evolves to be compact). This indicates that the iterative inference enables audio-visual representation to extract semantic discrimination from visual feature, and thus can reduce ambiguity in predicted sound separation masks.

\subsubsection{Effect of Visual Feature Extractor}
\begin{figure*}
	\centering
	\includegraphics[width=0.31\linewidth]{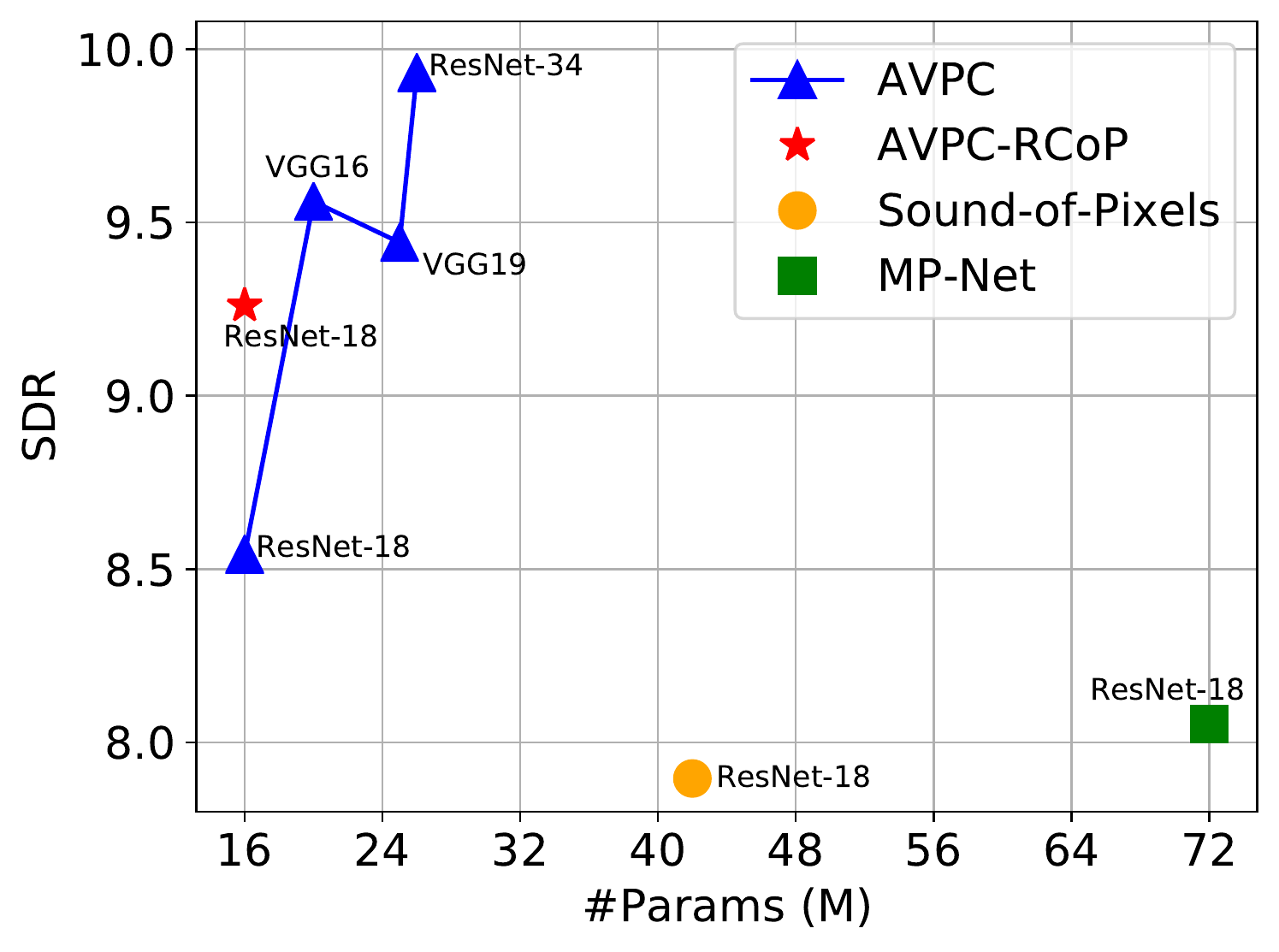}\hspace{1.3em}
	\includegraphics[width=0.31\linewidth]{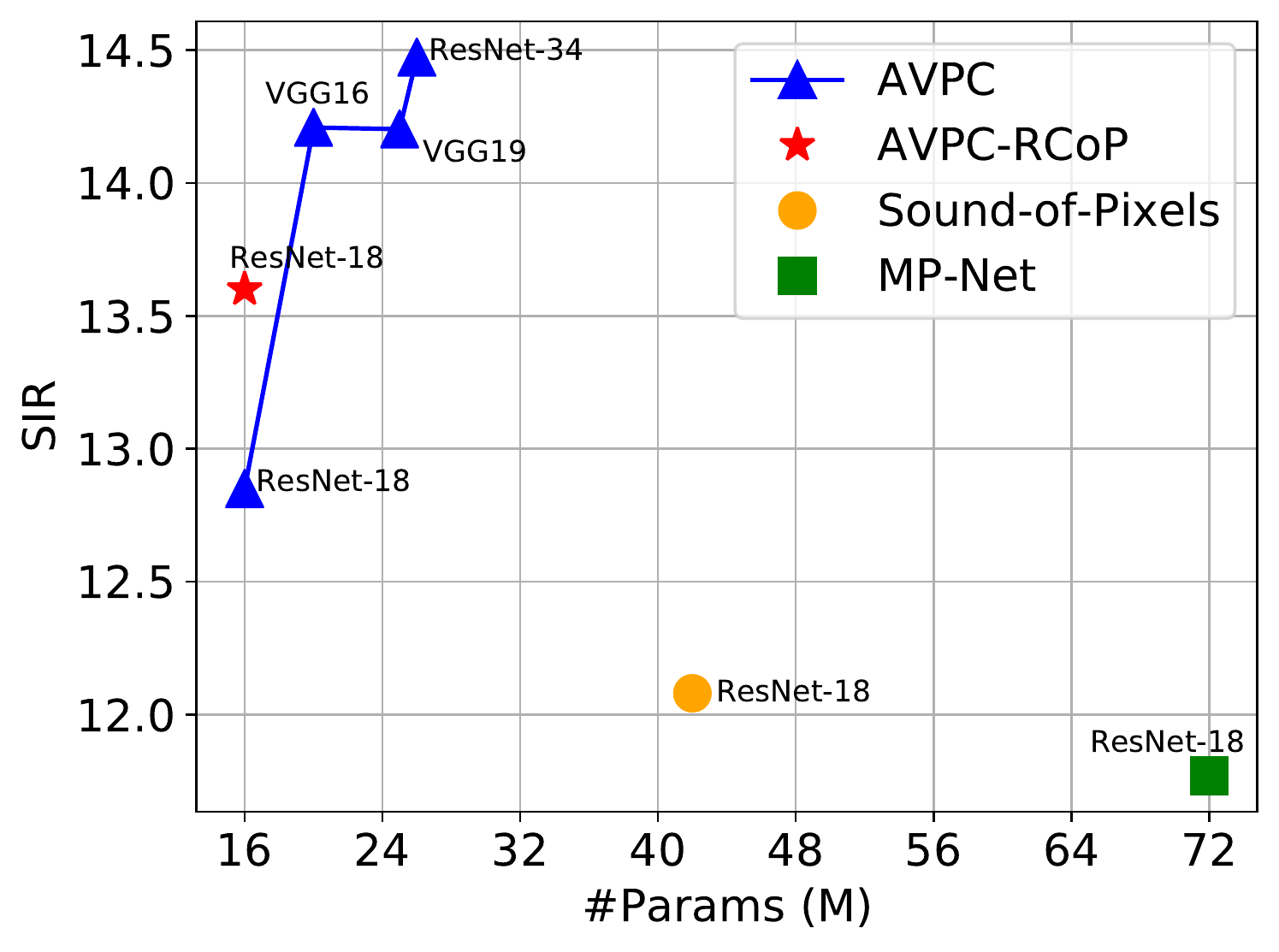}\hspace{1.3em}
	\includegraphics[width=0.31\linewidth]{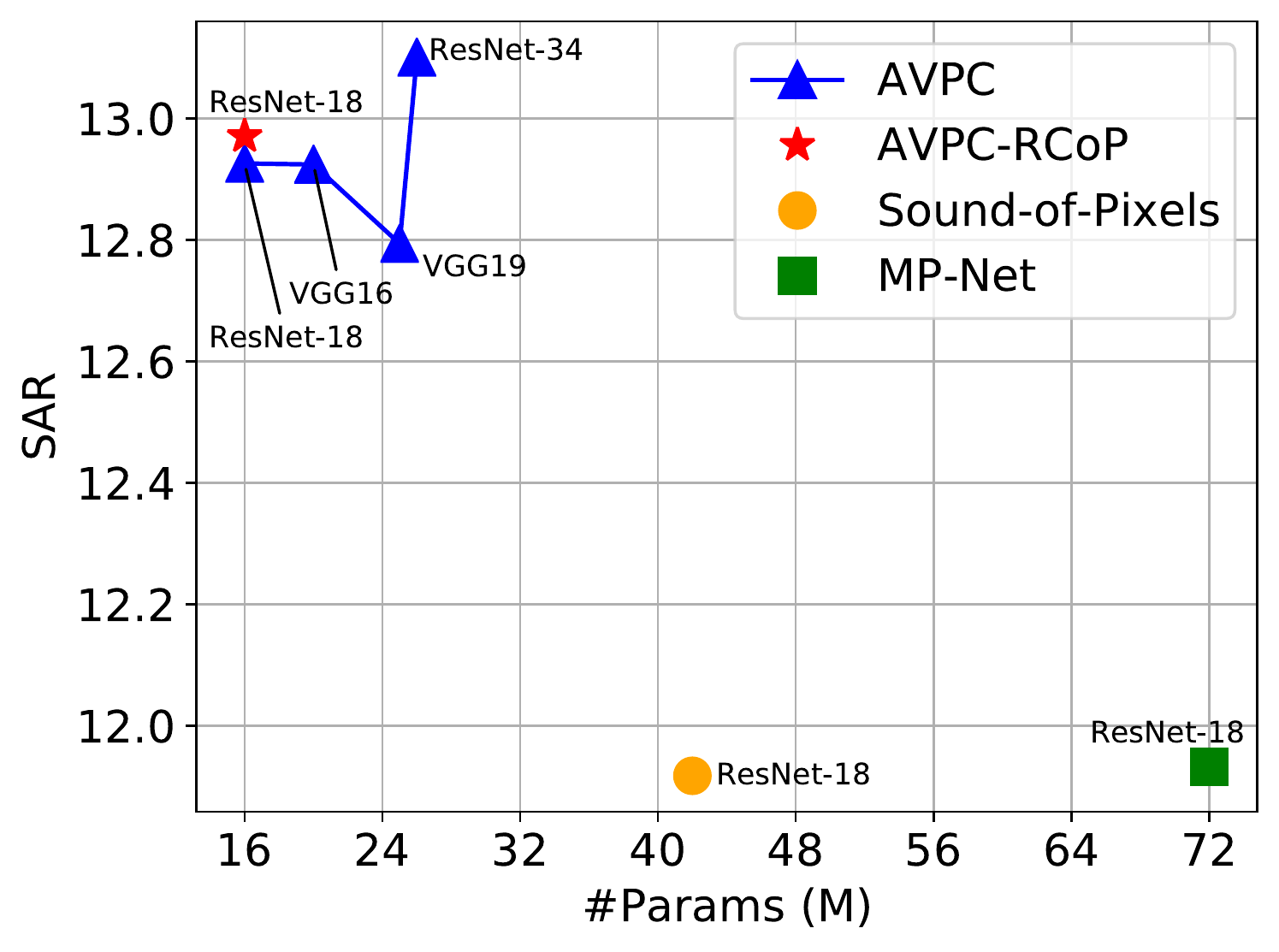}
	\caption{Performance comparison of visual sound separation methods using different visual backbone networks. The annotation next to each marker indicates the visual backbone network adopted in corresponding sound separation approach. Both Sound-of-Pixels and MP-Net employ U-Net to process the mixture sound spectrogram, while all AVPCs use PCNet instead.}
	\label{fig:perform_vs_num_params}
\end{figure*}

To analyze the influence of visual feature extractor used in AVPC, we report \emph{Performance vs. \#Params} trade-offs on MUSIC-11 test set in Fig. \ref{fig:perform_vs_num_params}. To this end, all AVPCs take PCNet with the same configuration to construct the sound separation network, and the difference among them is only in the used visual backbone networks. As shown in Fig. \ref{fig:perform_vs_num_params}, AVPCs equipped with different backbones consistently perform better and meanwhile have fewer parameters than two baseline methods, Sound-of-Pixels and MP-Net. Because in this comparison experiment the number of parameters of PCNet (4.69 M) is much smaller than that of U-Net (30.26 M), the AVPC taking ResNet-34 as visual backbone is still a parameter efficient method (26.27 M) compared with two baselines. These observations show the potential of AVPC to reduce memory cost on maintaining models. Additionally, with the increase of model size, the performance of AVPC is generally improved across three metrics. In particular, when employing ResNet-34 as the visual backbone network, AVPC achieves best sound separation quality on MUSIC-11 (SDR=9.93, SIR=14.47, SAR=13.10). All these results demonstrate that AVPC can gain from visual features that carry more discriminative cues.

\subsubsection{Effect of Visual Feature Map's Shape}
We conduct experiments to provide a sensitivity analysis about the shape of the visual feature map $\mathbf{f}_{n}$ (i.e., ablation on $\hat{\mathrm{H}}\times\hat{\mathrm{W}}$ while setting $K=16$). In Table \ref{tab:ablation_vis_fm_shape}, we find that compared with the smallest feature map ($1\times1$, as also used in Sound-of-Pixels and MP-Net) and the largest one ($7\times7$), the feature map with the shape of $2\times2$ benefits AVPC's sound separation performance most. This is presumably because large feature map not only conveys semantic discrimination related to sounding object, but also contains noise of background distractors; while small map leaves out too much information to serve as valid visual guidance. Therefore, we adopt the $2\times2$ visual feature map to guide sound separation in all cases.
\begin{table}[t]
	\renewcommand\arraystretch{1.2}
	\caption{Ablation on shape of the visual feature map. Results are reported on the held out MUSIC-11 test set ($N = 2$ mixture).}
	\label{tab:ablation_vis_fm_shape}
	\centering
	\begin{tabular}{l|S[detect-weight, table-format=1.2]S[detect-weight, table-format=2.2]S[detect-weight, table-format=2.2]S[detect-weight, table-format=2.2]}
		\hline
		$\hat{\mathrm{H}}\times\hat{\mathrm{W}}$    &{$1\times1$}    &{$2\times2$}    	 &{$4\times4$}      &{$7\times7$}  \\		   						
		\hline\hline
		{SDR $\uparrow$}    						&8.25            &\bfseries 8.54     &\uline{8.38}      &8.01        \\
		{SIR $\uparrow$}    						&12.65           &\uline{12.85}      &\bfseries 12.97   &12.42        \\
		{SAR $\uparrow$}    						&\uline{12.66}   &\bfseries 12.93    &12.64             &12.49        \\
		\hline
	\end{tabular}
\end{table}

\subsubsection{Effect of Feature Fusion Method}
\begin{table}[t]
	\renewcommand\arraystretch{1.2}
	\caption{Ablation on feature fusion methods. Results are reported on the held out MUSIC-11 test set ($N = 2$ mixture).}
	\label{tab:ablation_feature_fusion}
	\centering
	\begin{tabular}{l|S[table-format=2.2]|S[detect-weight, table-format=1.2]S[detect-weight, table-format=2.2]S[detect-weight, table-format=2.2]}
		\hline
		Fusion Method          			&{\#Params (M)}  	&{SDR $\uparrow$}    &{SIR $\uparrow$}    &{SAR $\uparrow$}  \\		   						
		\hline\hline
		Add          			        &65.85   	        &5.78                &11.13	   	          &9.70  	  		 \\
		Mul          			        &65.85   	        &6.76                &11.83	   	          &10.82  	  		 \\
		Cat          			        &70.05   	        &7.35                &12.51	   	          &10.97  	  		 \\
		Att          			        &99.44   	        &\uline{8.43}      	 &\uline{12.67}       &\uline{12.20}	 \\
		PCNet (Ours)				    &16.16   			&\bfseries 8.54  	 &\bfseries 12.85	  &\bfseries 12.93   \\
		\hline
	\end{tabular}
\end{table}
In AVPC, multimodal feature fusion is implicitly implemented through the alternative representation updating in PCNet. Here we compare the PCNet way with other four feature fusion methods: addition (Add), multiplication (Mul), concatenation (Cat) as widely used in \cite{Gao19,Hu20,Chatterjee21,Chatterjee22}, and attention-based modules (Att) \cite{Gan20,Zhu20a}. For fair comparisons, we leverage ResNet-18 as the default visual feature extractor in all compared methods, while use the 14-layer U-Net (the same as \cite{Gao19}) to process sound signal when the above four fusion methods are adopted. Besides, we set the attention module according to the configuration described in \cite{Gan20}. As we can see from Table \ref{tab:ablation_feature_fusion}, when fusing audio and visual features without employing parameterized modules, the three na\"ive methods (Add, Mul, and Cat) only achieve limited sound separation performance. While Att harvests results comparable to ours, it needs a heavily parameterized module for feature interaction. By contrast, our fusion with PCNet reaches better separation accuracy with smaller model size.

\subsection{Failure Cases}
\begin{figure}
	\centering
	\includegraphics[width=0.8\linewidth]{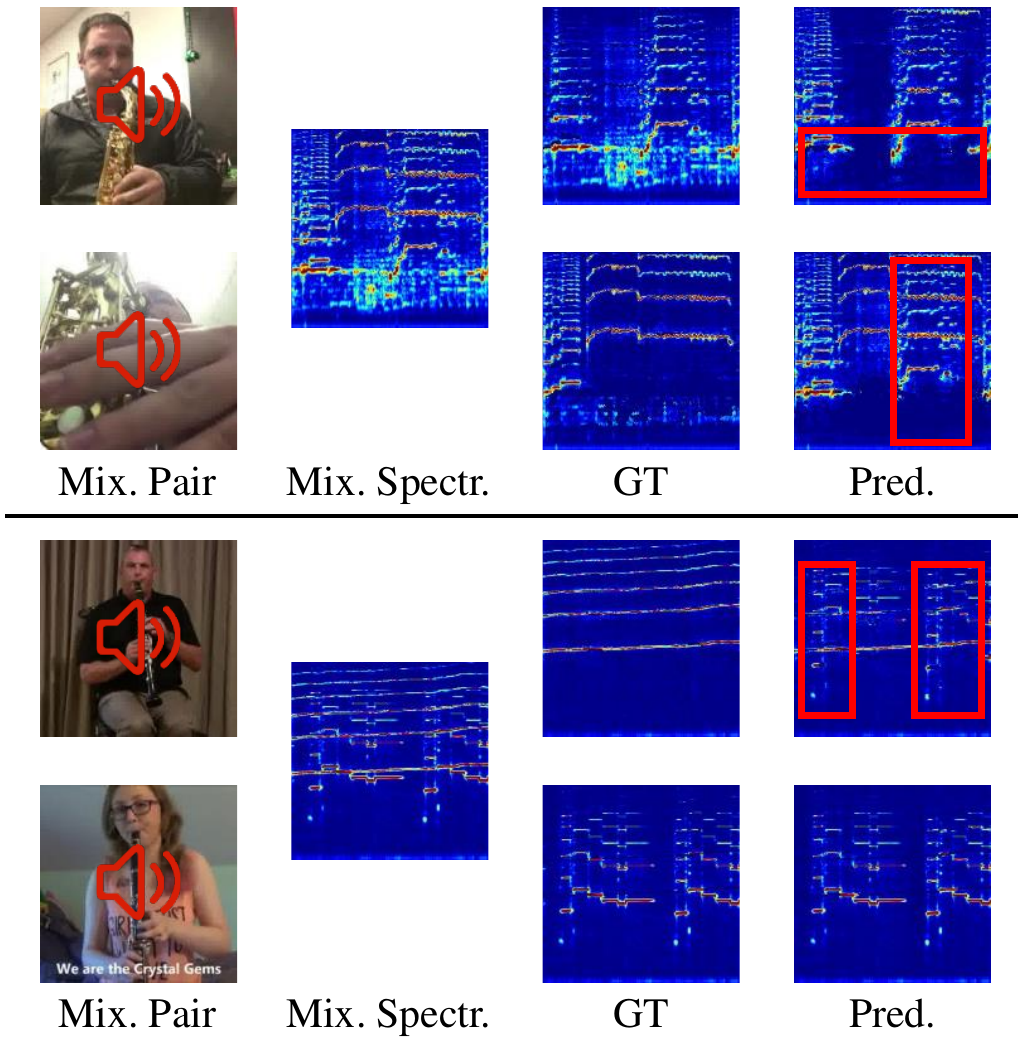}
	\caption{Failure cases of visual sound separation with AVPC. Homo-musical separation is a challenging task for our approach.}
	\label{fig:failure_cases}
\end{figure}
Fig. \ref{fig:failure_cases} illustrates failure cases of AVPC for homo-musical separation, which aims to separate sounds emitted by the same kind of musical instrument. The predicted spectrograms usually contain frequencies of the other sound component, as marked in red rectangles in Fig. \ref{fig:failure_cases}. This is mainly because AVPC only utilizes visual appearance cues of sounding object to guide separation, however, such visual cues are not distinguishable between the same category of instruments. To handle this challenging task, object motion-related cues, such as pixel-wise trajectories \cite{Zhao19} and human body keypoints \cite{Gan20}, have to be involved for visual feature extraction and integration.

\section{Conclusion and Future Work}\label{sec:conclusion}
We have presented a novel approach---audio-visual predictive coding (AVPC)---for visually-guided sound source separation. AVPC uses a simple video analysis network to derive semantic visual features, which then serve to guide the sound separation network to extract audio features, fuse multimodal information, and predict sound separation masks. Thanks to the parameter efficient network architecture and the iterative representation inference mechanism of predictive coding, AVPC not only achieves superior performance compared with conventional U-Net-based methods, but also significantly reduces the number of model parameters. Furthermore, we have introduced a self-supervised learning strategy RCoP that aims at learning two similar audio-visual representations of the target sound component, showing an effective way to further boost sound separation performance. Extensive experiments demonstrate the effectiveness of our approach on separating musical instrument sounds. 

We anticipate that this work would provide a new insight into multimodal representation learning. By viewing one modality feature with semantic discrimination as input and another modality feature with uncertainty as adjustable prior, one could leverage the cross-modal feature prediction in predictive coding to reduce such uncertainty. This mechanism has proven to be helpful to visual sound localization \cite{Song22}, and also possesses potential to benefit vision-language tasks (e.g., referring expression comprehension \cite{Kamath21}).

For visual sound separation, there exist open yet challenging problems that are not investigated in this work, such as separating sounds from sound mixtures with arbitrary numbers and types of sound components \cite{Xu19}, removing off-screen sound noise or separating intermittent sounds \cite{Zhu20a}, the homo-musical separation task \cite{Zhao19,Gan20}, \emph{etc}. In addition to develop task-oriented methods to mitigate these problems (e.g., from the perspective of disentangled representation learning \cite{Rouditchenko19,Song19,Song20,Chen23}), it is also necessary to collect high-quality video data that can cover diverse audio-visual scenarios \cite{Gao23}. This may be a direction of future investigation.

\ifCLASSOPTIONcaptionsoff
  \newpage
\fi

\bibliographystyle{IEEEtran}
\bibliography{IEEEabrv,mybibfile}

\end{document}